\documentclass [aps,prb,twocolumn,amssymb,amsmath,showpacs,floatfix,superscriptaddress]{revtex4-2}

\def\XXint#1#2#3{{\setbox0=\hbox{$#1{#2#3}{\int}$}
     \vcenter{\hbox{$#2#3$}}\kern-.5\wd0}}

\usepackage{amsmath}
\usepackage{amssymb}
\usepackage{amsfonts}
\usepackage{graphicx}
\usepackage{verbatim}
\usepackage{bm}
\usepackage{mathbbol}
\usepackage[dvipsnames]{xcolor}
\setcitestyle{compress}
\usepackage{mathtools}
\usepackage{esint}
\usepackage{tabularx}
\usepackage[normalem]{ulem}
\usepackage{graphicx,epsfig,amsfonts,amssymb}
\usepackage{bm}
\usepackage{times}
\usepackage{lipsum}
\usepackage{verbatim}
\usepackage{hyperref}
\hypersetup{
    colorlinks,
    citecolor=blue,
    filecolor=blue,
	    linkcolor=blue,
    urlcolor=blue
}
\usepackage{comment}

\newcommand{\beg}{\begin{equation}}
\newcommand{\en}{\end{equation}}

\newcommand{\overbar}[1]{\mkern 1.5mu\overline{\mkern-1.5mu#1\mkern-1.5mu}\mkern 1.5mu}
 
\begin{document}

\title{Trotter Transition in BCS Pairing Dynamics}

\author{Aniket Patra}
\affiliation{Center for Theoretical Physics of Complex Systems, Institute for Basic Science, Daejeon 34126, Republic of Korea}

\author{Emil A. Yuzbashyan}
\affiliation{Department of Physics and Astronomy, Center for Materials Theory, Rutgers University, Piscataway, NJ 08854, USA}
 
\author{Boris L. Altshuler}
\affiliation{Physics Department, Columbia University, 538 West 120th Street, New York, NY 10027, USA}

\author{Sergej Flach}
\affiliation{Center for Theoretical Physics of Complex Systems, Institute for Basic Science, Daejeon 34126, Republic of Korea}
\affiliation{Basic Science Program, Korea University of Science and Technology (UST), Daejeon, 34113, Republic of Korea}

\begin{abstract}

We study universal aspects of thermalization induced by Trotterization, a procedure routinely used in gate-based quantum computation. We use the reduced-BCS model --- quantum integrable with a classically integrable mean-field limit --- where the effects of Trotter chaos are expected to be particularly stark. The resulting Trotterized chaotic dynamics is characterized by its Lyapunov spectrum and rescaled Kolmogorov-Sinai entropy. The chaos quantifiers depend on the Trotterization time step $\tau$. We observe a Trotter transition at a finite step value $\tau_c \approx \sqrt{N}$. While the dynamics is weakly chaotic for time steps $\tau \ll \tau_c$, the regime of large Trotterization steps is characterized by short temporal correlations. We derive two different scaling laws for the two different regimes by numerically fitting the maximum Lyapunov exponent data. The scaling law of the large \(\tau\) limit agrees well with the one derived from the kicked top map. Beyond its relevance to current quantum computers, our work opens new directions -- such as probing observables like the Loschmidt echo, which lie beyond standard mean-field description -- across the Trotter transition we uncover.
 
\end{abstract}

\maketitle

Since the inception of von Neumann architecture, computing machines have employed time-discretized numerical methods to approximate solutions for ordinary differential equations \cite{Basic_Num1, Basic_Num2}, notably Hamilton's equations of motion. A similar philosophy is adopted in digital quantum simulation (DQS) \cite{DQS_Feynman, DQS_Lloyd, DQS_Nori, DQS_QuantH2} to discretize quantum time evolution operators using the Suzuki-Trotter decomposition \cite{Trot, Trot_Sympl_1, Trot_Sympl_2, Trot2, Trot3, Trot4, Trot5, Trot6, Trot7, Trot8, Trot9, Trot10, Trot11, Trot12}. In such algorithms, a.k.a.\ ``Trotterizations", a sharp transition in Trotter errors has been reported \cite{TrotterTrans1, TrotterTrans2, TrotterTrans3, TrotterTrans4, TrotterTrans5, TrotterTrans6, TrotterTrans7, TrotterTrans8, TrotterTrans9}. The very existence of a Trotter transition in strongly interacting quantum chaotic systems has subsequently been called into question \cite{AntiTrotterTrans}. In contrast to classical nonlinear dynamics, where the Lyapunov spectrum (LS) provides a universal measure of chaos \cite{Benettin, Skokos1, CSLS, Haenggi},  no comparable indicator exists for quantum many-body systems. Estimates of quantum chaos depend sensitively on the chosen observables \cite{Khinchin, Baldovin}, while finite system sizes further limit the reliability of such analyses. 

We investigate the onset of Trotter chaos through the LS of the reduced-BCS model \cite{BCS, BCS_Anderson1, BCS_Anderson2, BCS_Richardson, BCSUnivHam, BCSMFCorr, NonEqBCS1, NonEqBCS2, NonEqBCS3, NonEqBCS4, NonEqBCS5, NonEqBCS6}, which possesses a well-defined mean-field limit amenable to such analysis. Because both the quantum \cite{BCS_Richardson} and mean-field formulations in the thermodynamic limit \cite{BCSMFCorr, NonEqBCS1, NonEqBCS2} are integrable, this fine-tuned Hamiltonian offers complete analytical control and a fully accessible spectrum of solutions, providing a controlled setting in which the effects of Trotter chaos are expected to appear with particular sharpness. In this paper, we show that when the mean-field reduced-BCS dynamics are simulated using symplectic integrators, the system exhibits a Trotter transition --- from a weakly nonintegrable regime to a memoryless, fully ergodic one --- as the Trotter step size increases. Our approach relies on the equivalence between Trotterization and a class of classical symplectic integrators, in which the full Hamiltonian is divided into exactly solvable components and time evolution is constructed through successive applications of their individual propagators \cite{Neri, Yoshida, Koseleff1, McLachlan1, Koseleff2, McLachlan2, Laskar, Tao, Skokos2}. In implementing this scheme, we partition the Hamiltonian into two parts --- one containing no interactions and the other encompassing all-to-all interactions --- so that each part remains consistent with the mean-field approximation applied to the full BCS Hamiltonian.

In classical numerical simulations of integrable systems, chaos induced by time discretization is a computational artifact. Discretized integrators break integrability and can generate chaotic behavior even for finite step sizes. While such effects may be relevant to questions of long-time predictability and to certain esoteric yet fundamental issues in classical mechanics --- such as the possible relation between discretization-induced chaos, the KAM theorem, and the shadowing lemma --- they carry no direct physical significance. In digital quantum simulation (DQS), however, Trotterization constitutes the physical protocol for time evolution. As a result, the emergent Trotter chaos becomes an experimentally observable phenomenon that not only sets bounds on reliable simulation regimes and the class of observables measurable on quantum hardware, but also provides insight into how thermalization takes place in such devices --- insight that can be accessed through the classical-quantum correspondence, as demonstrated in this paper in the context of the reduced-BCS model. Moreover, the information encoded in the LS is universal, offering a general framework for characterizing thermalization and chaos across quantum platforms, and can even be exploited for state preparation --- for instance, enabling the initialization of fully chaotic, entangled states in the memoryless, ergodic Trotter regime.

We demonstrated that symplectic integrators introduce a hidden, time-dependent driving force, leading to chaos --- signaled by a positive maximum Lyapunov characteristic exponent (mLCE) --- in systems that are originally integrable, such as the Toda chain \cite{AP}. There, we were unable to analyze the scaling properties of the full LS, which are essential to understand the underlying thermalization mechanism. The simulations exhibited transient Floquet heating \cite{FloquetHeat_Rigol, FloquetHeat_Lazarides, FloquetHeat_Papic, FloquetHeat_Abanin1, FloquetHeat_Mori1, FloquetHeat_Mori2, FloquetHeat_Abanin2, FloquetHeat_Abanin3, FloquetHeat_Luitz}, which eventually led to numerical breakdowns due to the noncompact phase space and the exponential Toda potential. In contrast, the mean-field reduced-BCS model evolves on a compact phase space of fixed-length spins and contains no exponential interactions, thereby avoiding such instabilities and allowing a complete exploration of the post-Trotter transition regime. In this context, one also needs to mention prior studies that have examined how numerical integration schemes can induce chaos in integrable systems such as the sine-Gordon model, nonlinear Schr\"{o}dinger equations, and the spatially discrete Ablowitz-Ladik chains \cite{NumChaos1, NumChaos2, NumChaos3, NumChaos4, NumChaos5, NumChaos6, NumChaos7, NumChaos8}. However, these latter investigations often lack a systematic analysis of Lyapunov exponents, which is crucial to quantify the onset and degree of chaos.   

\begin{figure}
    \centering
    \includegraphics[trim={0cm 0cm 0cm 0cm},clip,scale=0.125]{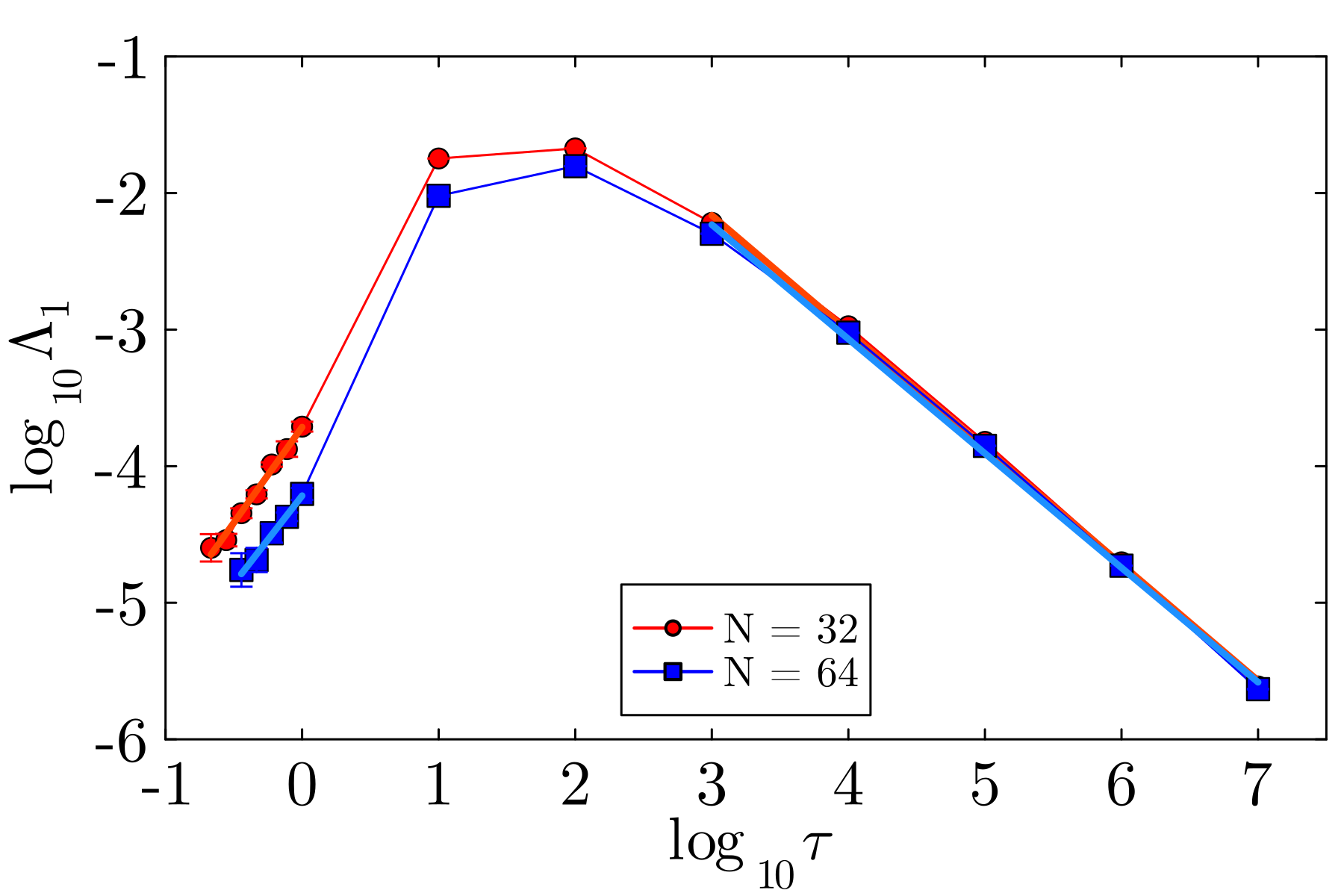}
    \caption{ We show $\log_{10} \Lambda_1$ as a function of $\log_{10} \tau$ for $N = 32$ and $64$. We have included the error bars. For a fixed $N$, we choose a configuration where all the spins point in random directions. The linear fits for $N = 32$ and $N=64$ to the first few points in the small $\tau$ regime are given by $y = 1.40x - 3.71$ and $y = 1.29x - 4.22$ respectively. On the other hand, the linear fit to the last few points in the large $\tau$ regime for both $N = 32$ and $N=64$ is given by $y = -0.85x + 0.39$. In the memoryless regime, the $N$ dependence of $\Lambda_1$ is indeed quite weak, see Fig.\ \ref{Fig:Large_tau_MLE}.}
    \label{Fig:Trott_Transition}
\end{figure} 

In prior instances \cite{TrotterTrans2, TrotterTrans3, TrotterTrans4, TrotterTrans5}, the Trotter transition has marked the onset of quantum chaos characterized by uncontrolled Trotter errors, aligning with predictions from random matrix theory (RMT). Additional clarity of knowledge about this dynamical phase transition is gained in models like the kicked top, where the presence of a well-defined classical limit helps bridge quantum complexity with classical intuition \cite{KickedTop_Haake, KickedTop_Greeks, KickedTop_Deutsch, KickedTop_Ghose}. The kicked top emerges during the Trotterization of the Lipkin-Meshkov-Glick model \cite{LMG}, an integrable system characterized by a single degree of freedom represented by the collective spin of $N$ spin-$1/2$ particles. Naturally, the kicked top dynamics can be analyzed using the established tools of single-body chaos theory. In contrast, the reduced-BCS model considered here is a many-body quantum model that can be realized in isolated mesoscopic grains \cite{BCSUnivHam}, or using ultracold atomic setups \cite{NonEqBCS5, NonEqBCS6}. In both the quantum and mean-field classical descriptions, strong interactions play a central role, making this model a natural testbed for exploring Trotter-induced thermalization in strongly interacting many-body systems. 

The two distinct Trotter regimes are marked by differing thermalization mechanisms. One approach to discern these mechanisms involves observing how typical observables relax over time to their ensemble-averaged values \cite{Khinchin}. However, identifying such suitable variables is nontrivial \cite{Baldovin}. Therefore, we use the mLCE and the LS as our primary diagnostic tools, which, in addition to being coordinate independent, remain invariant under a wide class of transformations \cite{Haenggi}. 

\begin{figure*}
    \centering
    \includegraphics[trim={0.5cm 0.25cm 0.0cm 0.0cm},clip,scale=0.85]{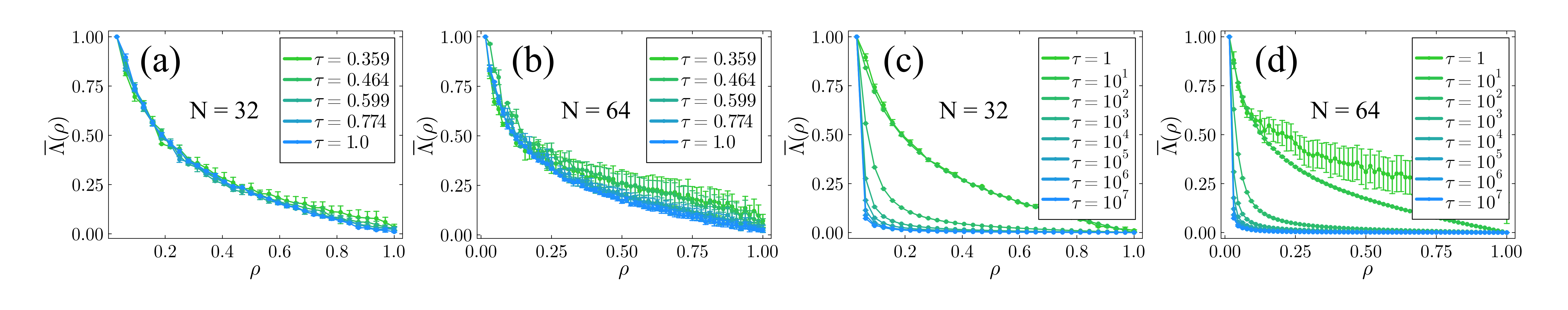}
    \caption{ Initiating from a fully random spin configuration where each spin points in a random direction, we show rescaled Lyapunov spectra for various step sizes $\tau$ for $N = 32$ in panels (a) and (c) and for $N = 64$ in panels (b) and (d). The small $\tau$ regime spectra in panels (a) and (b) $\overbar{\Lambda}(\rho) \equiv \Lambda_i/\Lambda_1$ are obtained for end time \(T_\mathrm{end} = 10^7\) and for step sizes \(\tau = 0.359 \text{ (limegreen)}, 0.464, 0.599, 0.774\) and \(1.0\) (dodgerblue). They show an approximate power law dependence on the normalized index $\rho \equiv i/N$ -- e.g., compare with the LRN spectra shown in Ref.\ \cite{Gabriel}. The large $\tau$ regime spectra in panels (c) and (d) $\overbar{\Lambda}(\rho)$ are obtained for step sizes \(\tau = 10^{0} \text{ (limegreen)}, 10^1, \ldots\) and \(10^7\) (dodgerblue) and for number of time steps \(N_\mathrm{steps} = 10^7\) and \(10^6\) respectively. They show a faster than an exponential decay as a function of the normalized index $\rho$. The transition from one regime to another takes place at $\tau_{c} \approx \sqrt{N}$. As seen in panels (c) and (d), the rescaled spectra remain similar to those of the small $\tau$ regime up to $\tau \approx 10$ but beyond this point their behavior changes abruptly.}
    \label{Fig:Rescaled_LS_Evolution}
\end{figure*} 

\begin{figure}
    \centering
    \includegraphics[trim={1.5cm 1cm 1cm 1cm},clip,scale=0.65]{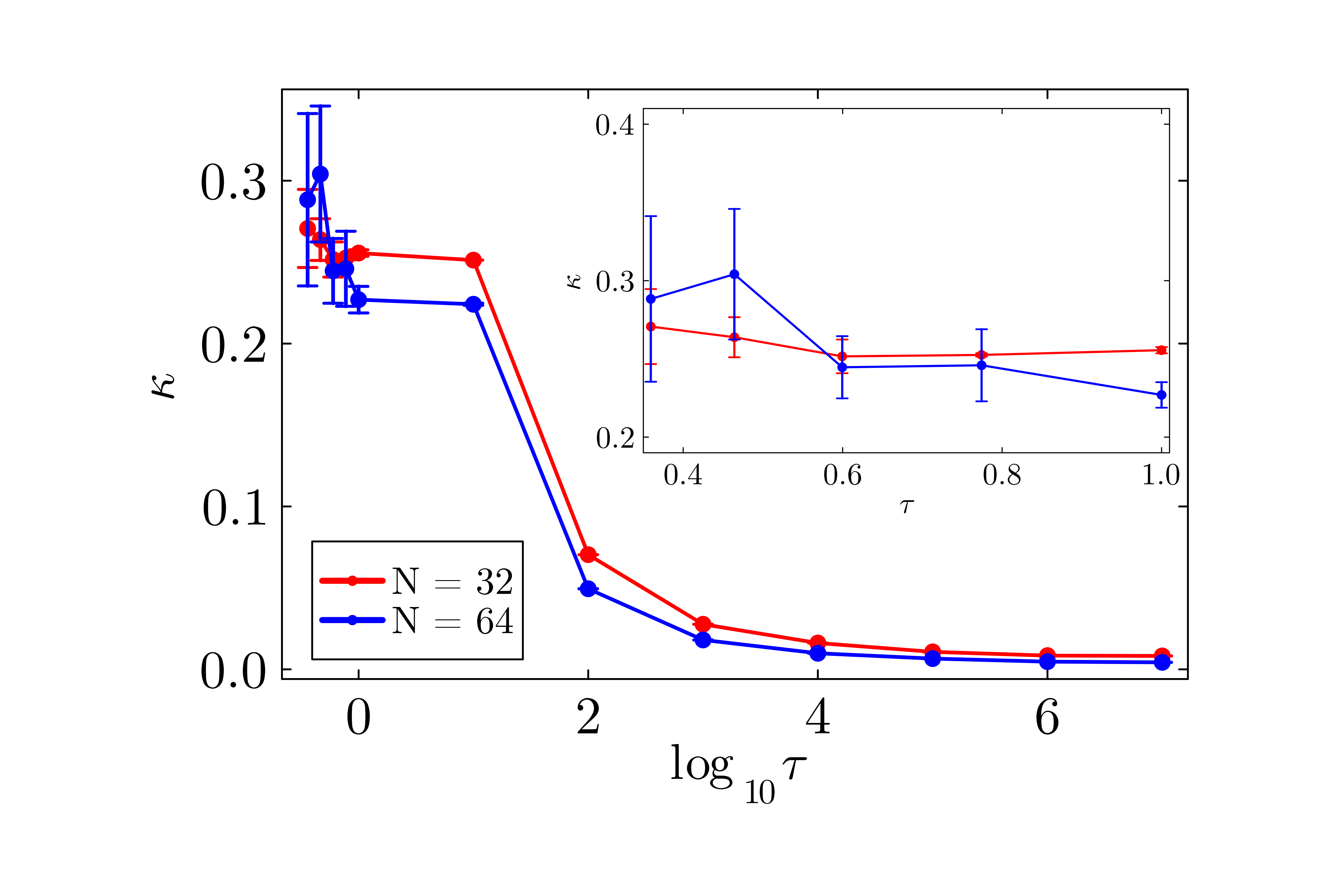}
    \caption{ We show the rescaled Kolmogorov-Sinai entropy \(\kappa\) as a function of $\log_{10} \tau$ for $N = 32$ and $64$. In the inset, we show a magnified \(\kappa\) versus \(\tau\) plot for the small \(\tau\) LRN regime.}
    \label{Fig:KS_Entropy}
\end{figure} 

To determine the LS \cite{Benettin, Skokos1, CSLS}, we start by evaluating the finite-time maximal Lyapunov characteristic exponent of order \( p \), denoted \( \Lambda^{p}(t) \). This exponent quantifies the exponential rate at which the volume of a \( p \)-dimensional parallelogram, formed by \( p \) linearly independent deviation vectors \( \bm{w}_1(t), \bm{w}_2(t), \ldots, \bm{w}_p(t) \), evolves over time. Mathematically, it is expressed as:
\[
\Lambda^{p}(t) = \frac{1}{t} \ln\left(\frac{\text{vol}_p[\bm{w}_1(t), \bm{w}_2(t), \ldots, \bm{w}_p(t)]}{\text{vol}_p[\bm{w}_1(0), \bm{w}_2(0), \ldots, \bm{w}_p(0)]}\right). \label{ft_pmLCE}
\]
Here, \( \text{vol}_p[\cdot] \) is the volume of the \( p \)-parallelogram spanned by the given vectors. The infinite-time limit of this quantity, \( \Lambda_1^{p} = \lim_{t \to \infty} \Lambda^{p}(t) \), provides the maximal Lyapunov characteristic exponent of order \( p \). To extract the individual Lyapunov characteristic exponents (LCEs), which collectively constitute the full LS, we utilize the relation:
\(\Lambda_p = \Lambda_1^{p} - \Lambda_1^{p-1}\), where \(\Lambda_1^{0} \equiv \Lambda_1\) is the mLCE, which is obtained using 
\begin{equation}
    \Lambda_1 = \lim_{k \to \infty}\frac{1}{k\tau} \sum_{j = 1}^{k} \ln \frac{ \lVert \bm{w}(j\tau) \rVert }{ \lVert \bm{w}( (j-1)\tau ) \rVert }. \label{Eq:mLCE}
\end{equation}
Here, \( \lVert \bm{w}(j\tau) \rVert \) is the magnitude of the deviation vector \( \bm{w} \) at time \( t = j\tau \), and \( \tau \) is the step size used to calculate all \( \Lambda_1^{p} \).  

In the small \( \tau \) regime, the integrability of the considered model is broken only slightly. Kolmogorov, Arnold, and Moser initiated the study -- now known as KAM theory -- of weakly perturbed integrable dynamics \cite{KAM11, KAM12, KAM13}. KAM theory predicts that certain invariant tori from the integrable dynamics survive and the dynamics remain quasiperiodic (i.e., near-integrable) if the perturbation to the integrable system is weaker than a critical strength \( \lessapprox \exp(-N\ln N) \) \cite{KAM11, KAM12, KAM13, KAM2, KAM3, KAM4, KAM_Perturb1, KAM_Perturb2, KAM_Perturb3}. Hence, one expects a perturbed integrable system with a large number of degrees of freedom to go out of this KAM regime very quickly. Nevertheless, such weakly nonintegrable macroscopic systems were observed to fall into two classes based on their mixing properties \cite{ClassHeisen, Mithun1, CarloGlass, Mithun2, Lubini, Mithun3, MerabPRL, MerabChaos, Gabriel, Weihua, Xiaodong}: (1) long-range network (LRN) and (2) short-range network (SRN). In LRN, the conserved actions of the corresponding integrable system undergo long-range coupling upon weak breaking of integrability, whereas the coupling is local in SRN. Our analysis indicates that the small \( \tau \) regime aligns with the characteristics of the LRN class.

We now describe the reduced-BCS Hamiltonian as:
\begin{equation}
    \hat{H}_\mathrm{BCS} = \sum_{j = 1}^{N} \varepsilon_j ( \hat{c}_{j\uparrow}^\dagger \hat{c}_{j\uparrow} + \hat{c}_{j\downarrow}^\dagger \hat{c}_{j\downarrow} ) - g \sum_{j,q = 1}^{N} \hat{c}_{j\uparrow}^\dagger \hat{c}_{j\downarrow}^\dagger \hat{c}_{q\downarrow} \hat{c}_{q\uparrow},   
    \label{Eq:H_BCS_Quant}
\end{equation}
which describes the Cooper pairing between time reversed single particle states \( |j\uparrow\rangle \) and \( |j\downarrow\rangle \) with energy $\varepsilon_j$ in finite-sized systems -- e.g., dirty superconductors \cite{BCS_Anderson2} and isolated metallic grains \cite{BCSUnivHam}. We consider equidistant single-particle energies $\varepsilon_j$ in the range $[-1/2, +1/2]$, and the coupling $g = 1/(N-1)$ to be the same as the single-particle level spacing -- as in \cite{BCSUnivHam}. This choice sets the bandwidth as the unit of energy, thereby fixing the dimensionless energy scale; correspondingly, all times --- including the Trotter step --- are measured in units of the inverse bandwidth. To obtain the mean-field approximation, this Hamiltonian \eqref{Eq:H_BCS_Quant} is first written using the Anderson pseudospin-1/2 operators: \( \hat{S}_j^{z} = ( \hat{c}_{j\uparrow}^\dagger \hat{c}_{j\uparrow} + \hat{c}_{j\downarrow}^\dagger \hat{c}_{j\downarrow} - 1 )/2 \) and \( \hat{S}_j^{-} = (\hat{S}_j^{+})^\dagger = \hat{c}_{j\downarrow} \hat{c}_{j\uparrow}. \) These pseudospins are defined on the unoccupied and the doubly occupied time reversed pairs corresponding to energy $\varepsilon_j$. The states where one of these states are occupied do not participate in the dynamics. In the limit \(N \to \infty\), we then replace the operators \( \hat{\bm{S}}_j \) by classical spins \( \bm{S}_j \) to obtain the following mean-field Hamiltonian \cite{BCSMFCorr, NonEqBCS1, NonEqBCS2, NonEqBCS3, NonEqBCS4}:  
\begin{equation}
    \label{Eq:H_BCS_MF}
    H_\mathrm{BCS} = \underbrace{    \sum_{j=1}^{N} 2\varepsilon_j S_j^{z}    }_{ H_\mathrm{free} }     \underbrace{    - g \sum_{j,k = 1}^{N} S_{j}^{+} S_{k}^{-}    }_{ H_\mathrm{int} }    \equiv    H_\mathrm{free}    +    H_\mathrm{int}.
\end{equation}

Solving the classical equations of motion for \(H_\mathrm{free}\) and \(H_\mathrm{int}\), we obtain 
\begin{equation}
\begin{split}
  e^{L_\mathrm{free}\tau}\bm{S}_j &=     \mathbb{R}_{\bm{z}} (  2  \varepsilon_j \tau   ) \cdot \bm{S}_j, \\
  e^{L_\mathrm{int}\tau}\bm{S}_j &=     \mathbb{R}_{\bm{z}} (  2  g J^{z}_0 \tau   ) \cdot \mathbb{R}_{\bm{n}_\mathrm{rot}} (  - 2  \lvert \bm{\Delta}_\mathrm{rot} \rvert \tau   ) \cdot \bm{S}_j,  
\end{split}
\label{Eq:Ind_Time_Evol}
\end{equation}
where \(L_\mathrm{free}\) and \(L_\mathrm{int}\) are the Liouvillian operators corresponding to \(H_\mathrm{free}\) and \(H_\mathrm{int}\) respectively \cite{Supp}. In Eq.\ \eqref{Eq:Ind_Time_Evol}, we designate an anticlockwise rotation matrix about a unit vector $\bm{n}$ by an angle $\theta$ by $\mathbb{R}_{\bm{n}}(\theta)$. The total spin components are written as \(J^\alpha = \sum_{j} S_j^\alpha\) for $\alpha = x, y,$ and $z$. Note that \(J^z\) is a constant of motion for $H_\mathrm{BCS}$, $H_\mathrm{free}$ and $H_\mathrm{int}$. We denote its initial value as $J^z_0$. We introduce the vector $\bm{\Delta} = gJ^x \bm{x} + gJ^y \bm{y}$ corresponding to the BCS order parameter \( \Delta = gJ^- = g(J^x - iJ^y) \). We note that \( \bm{\Delta}_\mathrm{rot} \) having the direction \(  \bm{n}_\mathrm{rot} \) defined as 
\begin{equation}
\begin{split}
  \bm{\Delta}_\mathrm{rot} = \bm{\Delta}(0) + g J^{z}_0 \bm{z}, \quad \bm{n}_\mathrm{rot} = \frac{\bm{\Delta}_\mathrm{rot}}{\lvert \bm{\Delta}_\mathrm{rot} \rvert},  
\end{split}
\label{Eq:BCS_Order_Param_in_RF}
\end{equation}
remains unchanged while we consider the evolution due to \( H_\mathrm{int} \). 

Using the above split of $H_\mathrm{BCS}$ in Eq.\ \eqref{Eq:H_BCS_MF} and the Baker–Campbell–Hausdorff formula, a symplectic integrator using Lie formalism approximates the operator \( e^{\tau L_\mathrm{BCS}} \) as
\begin{equation}
    \label{Eq:Gen_Sympl_Int}
    e^{\tau L_\mathrm{BCS}} = \prod_{j = 1}^{k} e^{a_j \tau L_\mathrm{free}}e^{b_j \tau L_\mathrm{int}} + 
    \mathcal{O}(\tau^p), 
\end{equation}
with \( \sum_{j} a_j = \sum_{j} b_j = 1 \), where \(L_\mathrm{BCS}\) is the Liouvillian operator corresponding to \(H_\mathrm{BCS}\). The precision and order \( (p-1) \) of the integrator depend on $k$ and on the specific values of $\{a_j\}$ and $\{b_j\}$. In our numerics, we use the 2nd order $\mathrm{SABA}_2$ integrator, for which \( p = 3 \) \cite{Laskar, Supp}. 

\begin{figure}
    \centering
    \includegraphics[trim={0cm 0cm 0cm 0cm},clip,scale=0.125]{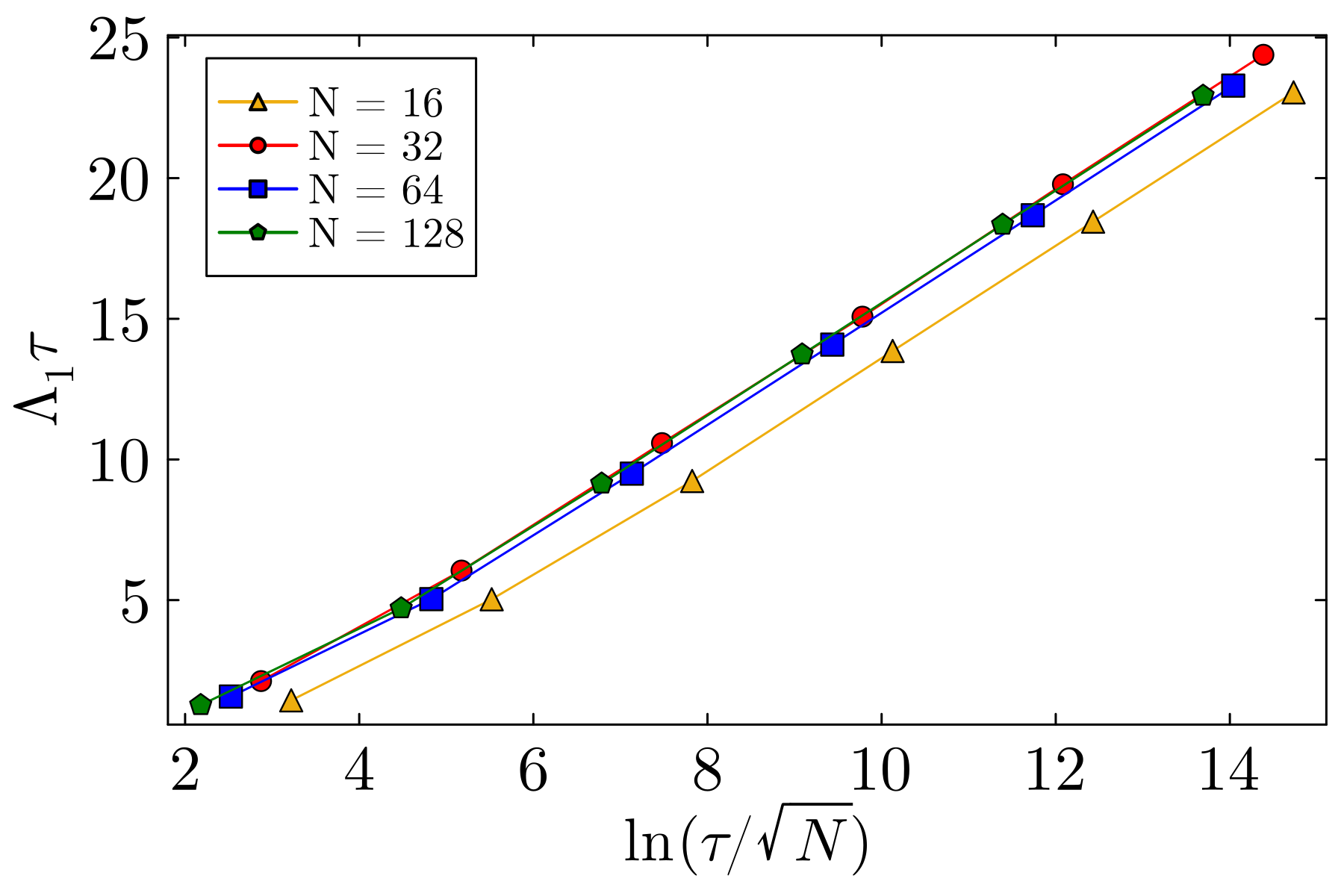}
    \caption{ We show $\tau \Lambda_1$ (obtained from numerics) as a function of $\ln (\tau/\sqrt{N})$ for $N = 16, \; 32,\; 64$ and $128$ with completely random initial spin configurations. The slope of the linear fit to the $N = 16$ data is $1.96 \pm 0.02$, whereas the slope for the $N = 32,\; 64$ and $128$ data is $1.99 \pm 0.01$. }
    \label{Fig:Memoryless_Scaling}
\end{figure}

We establish the Trotter transition -- e.g., in Fig.\ \ref{Fig:Trott_Transition} --  using typical spin configurations where each spin is randomly oriented. To compute \(\Lambda_1\) for the reduced-BCS model, we use \(3N\)-dimensional deviation vectors \( \bm{w}(t)= \left[\delta \bm{S}_1 (t), \ldots, \delta \bm{S}_N (t)
\right] \) in Eq.\ \eqref{Eq:mLCE}. We extract a power law for the LRN regime: \(\Lambda_1 \propto \tau^\eta\) where $\eta = 1.40 \pm 0.06$ for $N = 32$ and $\eta = 1.29 \pm 0.09$ for $N = 64$. This exponent being very close to the exponent \((\eta = 1.36)\) obtained for the Toda chain dynamics \cite{AP} -- also simulated using the \(\mathrm{SABA}_2\) integrator -- strongly hints at a universal physics. In the memoryless regime, we obtain \(\eta = -0.85 \pm 0.02\) both for $N = 32$ and $N = 64$. Later we obtain a more accurate scaling for \(\Lambda_1\) by comparing the memoryless large \(\tau\) reduced-BCS dynamics with the kicked top dynamics -- cf. Fig.\ \ref{Fig:Memoryless_Scaling}.         

We study how the scaling of the LS with \(\tau\) changes across the Trotter transition in Fig.\ \ref{Fig:Rescaled_LS_Evolution} using the completely random initial spin configurations of Fig.\ \ref{Fig:Trott_Transition}. Since \(\lVert \bm{S}_i (t) \rVert = 1/2\) for all \(i\) at any \(t\), \(N\) LCE values \(\{ \Lambda_N, \Lambda_{N+1}, \ldots, \Lambda_{2N} \}\) always remain equal to zero. Symplecticity ensures that each positive Lyapunov exponent \( \Lambda_i \) has a corresponding negative exponent \( \Lambda_{3N - i +1} = - \Lambda_i \), reflecting the system's phase-space volume preservation. As a result, it is sufficient to analyze the rescaled positive LCEs \(\overbar{\Lambda}_{i} \equiv \Lambda_i/\Lambda_1\) to capture the characteristics of the Trotter transition in Fig.\ \ref{Fig:Rescaled_LS_Evolution}, where we show how the spectra evolve from the weakly chaotic regime to the large $\tau$ regime. In the small \( \tau \) regime, the rescaled LS exhibits a power-law decay, consistent with prior findings -- cf. \cite{Gabriel}. Conversely, in the memoryless regime, the decay of \( \overline{\Lambda}_i \) is sharper than exponential. The normalized spectrum \( \overline{\Lambda}(\rho) \) versus \( \rho = i/N \) appears to reach saturation in both regimes.

To classify the two Trotter phases, we obtain the rescaled Kolmogorov-Sinai (KS) entropy
\begin{equation}
    \label{Eq:KS_Entropy}
    \kappa = \frac{1}{N-1}\sum_{i = 2}^{N} \overbar{\Lambda}_{i} = \int_{0}^{1} \overbar{\Lambda}(\rho)\, d\rho, 
\end{equation}
from the LS. In the LRN regime, \(\kappa\) saturates to a positive value close to \(\kappa_\mathrm{LRN} \approx 0.3\) in Fig.\ \ref{Fig:KS_Entropy}. With the increase of \(\tau\), the KS entropy saturates to a very small value \( \kappa_\mathrm{ML} \ll \kappa_\mathrm{LRN}\) in the memoryless regime.

\begin{figure}
    \centering
    \includegraphics[trim={2.5cm 1.5cm 2.5cm 1.5cm},clip,scale=0.40]{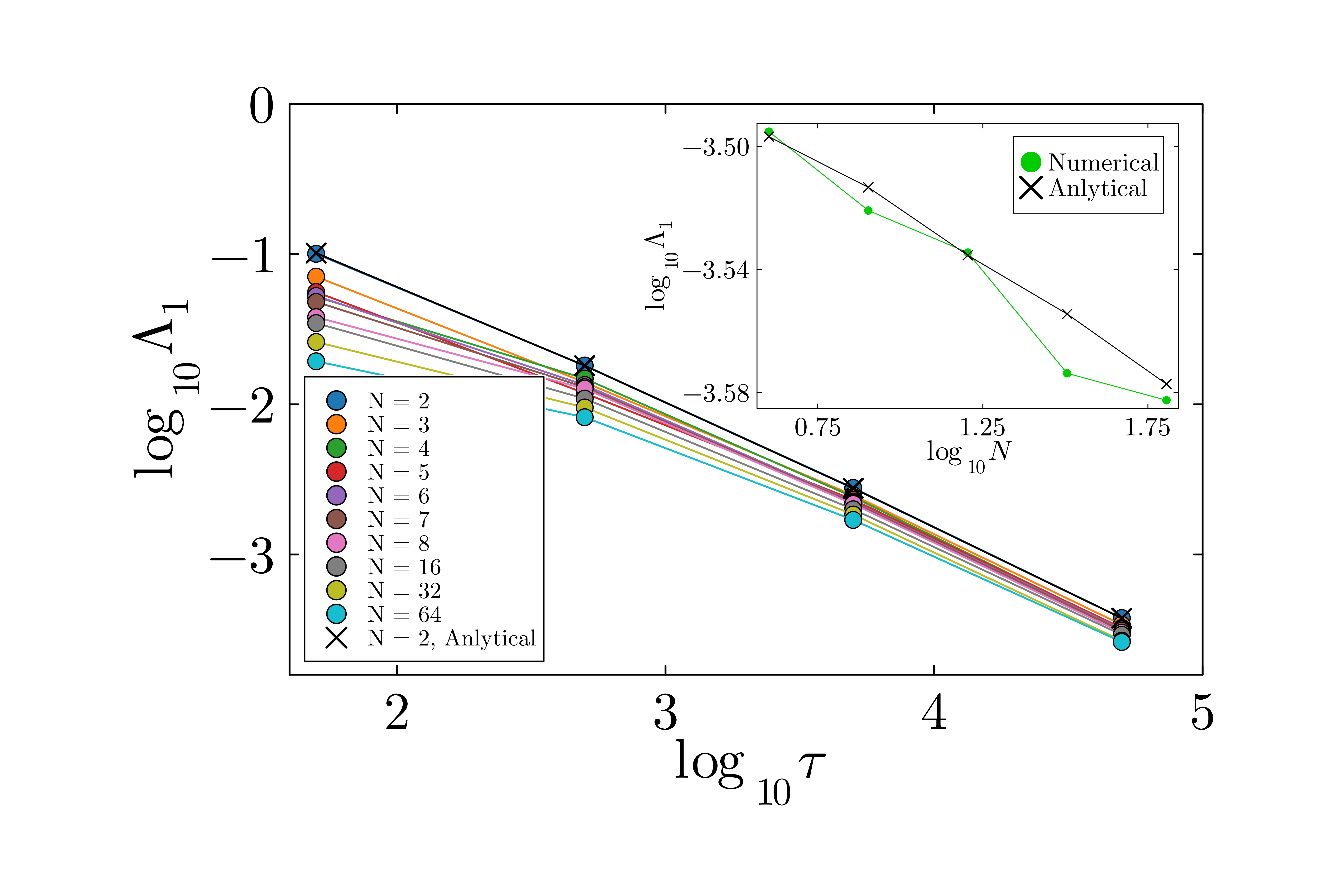}
    \caption{ We show the $\log$-$\log$ plot of $\Lambda_1$ versus $\tau$ for $N = 2, 3, 4, 5, 6, 7, 8, 16, 32$ and $64$ with $50 \leq \tau \leq 5 \times 10^4$. The numerical values of $\Lambda_1$ (dark blue circles) for $N = 2$ at different $\tau$ values coincide with the data (black cross) obtained from our semi-analytic method requiring only ensemble averaging and no time averaging for $N = 2$ with particle-hole symmetric initial condition. In the inset, we show the $\log$-$\log$ plot of $\Lambda_1$ versus $N$ for $\tau = 5\times 10^4$. Here lime-green circles represent values obtained from our numerical calculation (linear fit: $y = -0.076x - 3.45$), whereas black crosses indicate results from our semi-analytic method (linear fit: $y = -0.067x - 3.45$). }
    \label{Fig:Large_tau_MLE}
\end{figure}

Although integrability is broken by our symplectic integration scheme, it still conserves \(J_0^{z}\). In our model, Trotter chaos depends only weakly on \(J_0^{z}\) for small \(\tau\), and this dependence weakens further in the memoryless regime \cite{Supp}. To show this, we first determine the energy extrema for a given \(J_0^{z}\) -- see the \(E(J_0^{z})\) spectrum in \cite{Supp}. Given the rotational symmetry of the BCS Hamiltonian \eqref{Eq:H_BCS_MF} about the \(z\)-axis, we can, without loss of generality, consider extremal spin configurations to lie in the $xz$-plane. The energy maxima are obtained by making the first $N/2 - J_0^z$ spins point down and the rest pointing up. In the minimum \(E\) configuration, the spin with the lowest single-particle energy exhibits the largest positive $z$-component; subsequent spins progressively tilt, resulting in the spin with the highest single-particle energy possessing the largest negative $z$-component. Completely random spin configurations with \(E \approx 0, J_0^{z} \approx 0,\) and \(\Delta \approx \mathcal{O}(1/\sqrt{N})\) also lie in the middle of this \(E(J_0^{z})\) spectrum. The MLE is maximum for initial spin configurations that lie in the middle of the \(E(J_0^{z})\) spectrum. 

In the following, we analyze the dependence of \(\Lambda_1\) on \(\tau\) in the large \(\tau\) regime. Beyond offering insight into thermalization processes relevant to quantum computation, this analysis identifies the step size at which the Trotter transition occurs. The large \(\tau\) behavior is captured by the kicked top map \cite{KickedTop_Haake, KickedTop_Greeks, KickedTop_Deutsch, KickedTop_Ghose}, from which the analytical scaling of \(\Lambda_1\) is obtained. This scaling provides a key criterion for determining the critical step size \(\tau_c\): its breakdown signals the onset of the Trotter transition.

In Fig.\ \ref{Fig:Large_tau_MLE}, we show a $\log$-$\log$ plot of $\Lambda_1$ versus $\tau \geq 50$ for several values of $N \geq 2$. Here, in the initial configuration, the spins point in random directions on the \(xy\)-plane and \(J_0^z = 0\). These curves exhibit minimal dependence on \(N\) and are independent of initial conditions. Therefore, we consider the system with two spins and particle-hole symmetry \cite{Supp}. Due to symmetry, the interacting system can be effectively described by a single spin. Based on our numerical simulations, the mLCE reaches saturation after just one time step \cite{Supp}. In contrast, within the LRN regime, the mLCE curve requires numerous time steps to reach saturation, reflecting the system's persistent memory effects. This behavior justifies referring to the large \(\tau\) strongly chaotic regime as ``memoryless." Using this and the ergodic hypothesis \cite{KickedTop_Greeks}, we calculate the mLCE for the $N = 2$ particle-hole symmetric case from the Jacobian of the \(\mathrm{SABA}_2\) map. We obtain \(\Lambda_1\) for large \(N\) by repeating the same calculation for the new coupling constant \(g_{N} = 1/\sqrt{13N/3}\) in \(H_\mathrm{int}\) -- cf. the inset of Fig.\ \ref{Fig:Large_tau_MLE} and \cite{Supp}. Analytically determining the mLCE for a dynamical system is generally challenging. However, in the memoryless regime characterized by strong global chaos, an analytical scaling becomes feasible. This exception underscores the unique nature of this regime.      

Applying the \(\mathrm{SABA}_2\) map repeatedly for the $N = 2$ particle-hole symmetric system is approximately equivalent to applying the kicked top map \cite{KickedTop_Haake, KickedTop_Greeks, KickedTop_Deutsch, KickedTop_Ghose}: 
\begin{equation}
    \label{Eq:Kicked_Top_H}
    H = A(S^x)^2 + BS^z\sum_{n} \delta(t-n),
\end{equation}
twice within the same time step with anisotropy parameter \(A = 4 b_1 g\tau\). The first and second kicks correspond to the magnetic field strengths \(B = 2a_1\tau \bmod \pi/2\) and \(B = a_2\tau \bmod \pi/2\), respectively. The values of \(\{a_1, a_2, b_1\}\) for the \(\mathrm{SABA}_2\) integrator are provided in \cite{Supp}. Edge effects in the initial and final time steps have negligible influence on the scaling behavior of the mLCE. For \(A \gg 1\), using the approach of Refs.\ \cite{KickedTop_Greeks} and \cite{KickedTop_Deutsch}, and noting that \(g = \delta = 1\) for \(N = 2\), we find  
\begin{equation}
    \label{Eq:Kicked_Top_mLCE_N=2}
    \tau\Lambda_1 \approx \ln\big[ (2\tau)^2 \lvert 
    \sin(2a_1\tau) \sin(a_2\tau) \rvert \big] + C_2,
\end{equation}
where \(C_2\) is a \(\tau\) independent constant. This indeed agrees with the mLCE obtained from the numerics in Fig.\ \ref{Fig:Large_tau_MLE} \cite{Supp}.  

This analysis can be generalized for large \(N\) with \(J_0^z = 0\), where we have \( \lvert \bm{\Delta}_\mathrm{rot} \rvert \propto 1/\sqrt{N} \) and the anisotropy parameter \(A \sim 2b_1\tau/\sqrt{N}\). Using Eq.\ \eqref{Eq:Kicked_Top_mLCE_N=2} and neglecting the subleading terms, we obtain 
\begin{equation}
    \label{Eq:Kicked_Top_mLCE_Gen_N}
    \tau\Lambda_1 \approx 2\ln\Big( \tau/\sqrt{N} \Big) + C_N,
\end{equation}
where \(C_N\) is a \(\tau\) independent constant. We verify this scaling in Fig.\ \ref{Fig:Memoryless_Scaling}. This scaling \eqref{Eq:Kicked_Top_mLCE_Gen_N} explains the approximate power law \(\eta = -0.85 \pm 0.02\) and the weak \(N\) dependence in the memoryless regime. Recognizing that the scaling applies when the anisotropy parameter \(A = \tau/\sqrt{N} \gg 1\), we estimate the Trotter transition point as \(\tau_c \approx \sqrt{N}\). This aligns with our numerical findings for \(N = 32\) and \(64\) in Figs.\ \ref{Fig:Trott_Transition} and \ref{Fig:KS_Entropy}. Recall that the step size is measured in inverse bandwidth. If the bandwidth is rescaled from $1$ to $\alpha$ while keeping $N$ fixed, the single-particle energies transform as $\varepsilon_j \to \alpha \varepsilon_j$. Upon simultaneously rescaling the coupling $g \to \alpha g$, the critical step size for the Trotter transition should scales as $\sqrt{N}/\alpha$.  

In conclusion, we examined the Trotterized BCS dynamics, identifying a Trotter transition via Lyapunov spectra. For small Trotter steps, the system exhibits weakly nonintegrable long-range network behavior. In contrast, large steps lead to a memoryless regime characterized by strong global chaos, where analytical scaling becomes tractable. These findings suggest several research directions. The exactness of mean-field theory for local observables in the reduced-BCS model \cite{NonEqBCS4} implies potential realizability on quantum computers, particularly those with all-to-all qubit connectivity \cite{DQS_QuantH2}. Owing to this classical–quantum correspondence, the Lyapunov spectrum analysis is directly applicable to understanding thermalization processes on quantum hardware. Furthermore, its universality makes the approach broadly relevant across different quantum platforms. Additionally, simulating such symplectic dynamics in ultracold atom setups could provide further insights \cite{NonEqBCS5, NonEqBCS6}. Future investigations might explore the dependence of the emergent power-law exponent \(\eta\) in LRN on different integrable and nonintegrable models, as well as on various symplectic integrators. Studying the effects of disorder and noise on the Trotter transition could also be illuminating. Moreover, examining nonlocal quantities -- such as entanglement entropy and the Loschmidt echo -- for which the standard BCS mean-field theory breaks down, may reveal additional aspects of the transition in the \textsl{quantum} dynamics of the BCS model.

AP and SF acknowledge the financial support from the Institute for Basic Science (IBS) in the Republic of Korea through Project No. IBS-R024-D1. AP also thanks Miguel de Jes\'{u}s Gonz\'{a}lez Mart\'{i}nez for several illuminating discussions.

\clearpage
\begin{center}
\textbf{\large Supplemental Material}
\end{center}

\setcounter{equation}{0}
\setcounter{figure}{0}
\setcounter{table}{0}
\setcounter{section}{0}
\makeatletter
\renewcommand{\theequation}{S\arabic{equation}}
\renewcommand{\thefigure}{S\arabic{figure}}
\renewcommand{\thetable}{S\arabic{table}}
\renewcommand{\bibnumfmt}[1]{[S#1]}
\renewcommand{\citenumfont}[1]{S#1}
\renewcommand{\thesection}{S\arabic{section}}
\renewcommand{\theHtable}{S\thetable}
\renewcommand{\theHfigure}{S\thefigure}
\renewcommand{\theHequation}{S\theequation}

\section{Split-Step Dynamics of the BCS Hamiltonian}
\label{Sec:SABA2_BCS_Soln}

In this section, we describe how to integrate the mean-field BCS Hamiltonian using a symplectic integrator scheme, which performs canonical transformations at each time step \cite{Yoshida_sup, Laskar_sup, Tao_sup, Skokos2_sup}. The structure of the Hamiltonian naturally divides into two distinct components
\begin{equation}
    \label{SEq:H_BCS}
    H_\mathrm{BCS} = \underbrace{    \sum_{j=1}^{N} 2\varepsilon_j S_j^{z}    }_{ H_\mathrm{free} }     \underbrace{    - g \sum_{j,k = 1}^{N} S_{j}^{+} S_{k}^{-}    }_{ H_\mathrm{int} }    \equiv    H_\mathrm{free}    +    H_\mathrm{int},
\end{equation}
where the norms of the individual spins $\bm{S}_j$ are constants of motion, and we set it to $1/2$. We have chosen equidistant single-particle energies $\varepsilon_j$ in the range $[-1/2, +1/2]$. We also have $g = \delta$, where $\delta = 1/(N-1) \approx 1/N$ is the single-particle level spacing. We now obtain the exact time evolution of the spins and their deviations for each Hamiltonian component for a single time step. We use this to implement the tangent map method \cite{Skokos1_sup} to calculate Lyapunov spectra in various situations \cite{Benettin_sup, CSLS_sup}.

First, we write the Hamilton's equations of motion and the variational equations for each spin, governed by the free and the interacting Hamiltonian components as
\begin{subequations}
\begin{equation}
    L_\mathrm{free}\colon \begin{cases}
    \frac{    d    \bm{S}_j    }{    dt    }    =     \left\lbrace     \bm{S}_j,      H_\mathrm{free}     \right\rbrace    =    2 \varepsilon_j \bm{z} \times \bm{S}_j,   \\[10pt]
    \frac{    d    \delta\bm{S}_j    }{    dt    } = 2 \varepsilon_j \bm{z} \times \delta\bm{S}_j, 
\end{cases}
\label{SEq:Tan_Map_HFree}
\end{equation}
\begin{equation}
L_\mathrm{int}\colon \begin{cases}
    \frac{    d    \bm{S}_j    }{    dt    }    =     \left\lbrace     \bm{S}_j,      H_\mathrm{int}     \right\rbrace    =    -2 \bm{\Delta} \times \bm{S}_j,   \\[10pt]
    \frac{    d    \delta\bm{S}_j    }{    dt    } = - 2\delta\bm{\Delta} \times \bm{S}_j - 2\bm{\Delta} \times \delta\bm{S}_j, 
\end{cases}
\label{SEq:Tan_Map_HInt}
\end{equation}
\end{subequations}
where $L_\mathrm{free}$ and $L_\mathrm{int}$ are the Liouvillian operators defined via the fundamental Poisson bracket for spins \cite{AidanBCS_sup} 
\begin{equation}
    \left\lbrace S^{a}_j,  S^{b}_k \right\rbrace     =     \delta_{jk} \varepsilon^{abc} S^{c}_j.
\label{SEq:PB_AM}
\end{equation}   
Here we have introduced the BCS order parameter $\Delta$ and the corresponding BCS order parameter vector $\bm{\Delta}$   
\begin{equation}
    \Delta   =     g    \sum_{k=1}^{N}    S_{k}^{-}     =     \Delta_x - i \Delta_y, \; \textrm{with} \;\;    \bm{\Delta}    =    \Delta_x \bm{x}    +    \Delta_y \bm{y},      
\label{SEq:BCS_OP}
\end{equation}
where $S_{j}^{\pm} = S_j^{x} \pm i S_j^{y}$. We also note that $J^z = \sum_{j} S_j^z$ is a constant of motion for $H_\mathrm{BCS}$, $H_\mathrm{free}$ and $H_\mathrm{int}$. Since the total $z$-component value at any time is equal to its initial value, we denote it as $J^z_0$.  

\subsection{Solving the Dynamics of $H_\mathrm{free}$}
\label{Sec:Free_Part_BCS_Soln}

The Hamilton equations and the variational equations due to $H_\mathrm{free}$ are identical to the Bloch equation with a constant magnetic field $-2 \varepsilon_j \bm{z}$. As a result, we write the solutions \( (\bm{S}_j^\prime, \delta\bm{S}_j^\prime) \) after one \(\tau\) time step explicitly as
\begin{equation}
e^{L_\mathrm{free}\tau}\colon \begin{cases}
    \bm{S}_j^\prime    =     \mathbb{R}_{\bm{z}} (  2  \varepsilon_j \tau   ) \cdot \bm{S}_j, \\[5pt]
    \delta\bm{S}_j^\prime = \mathbb{R}_{\bm{z}} (  2  \varepsilon_j \tau   ) \cdot \delta\bm{S}_j, 
\end{cases}
\label{SEq:Tan_Map_Soln_HFree}
\end{equation}
where $\mathbb{R}_{\bm{z}}(\theta)$ denotes an anticlockwise rotation matrix about the $z$-axis by an angle $\theta$, and $\tau$ is the integrator time step. 

\subsection{Solving the Dynamics of $H_\mathrm{int}$}
\label{Sec:Interacting_Part_BCS_Soln}

The equations of motion \eqref{SEq:Tan_Map_HInt} for $H_\mathrm{int}$ are more complicated. The magnetic field in the Bloch equations depends on the spins themselves, which makes them nonlinear. Summing the equations of motions for the $x$ and $y$-components over $j$, we obtain that the BCS order parameter vector $\bm\Delta$ obeys the equation
\begin{equation}
    \frac{    d    \bm{\Delta}    }{    dt    }    =    2  g J^{z}_0 \bm{z} \times \bm{\Delta}.
\label{SEq:EOM_Del_Class_HInt_Vector}
\end{equation} 
This means that $\bm{\Delta}$ simply rotates in an anticlockwise direction with an angular velocity $2  g J^{z}_0$. In terms of $\Delta   =     g    \sum_{k=1}^{N}    S_{k}^{-}     =     \Delta_x - i \Delta_y$, we write the solution of Eq.\ \eqref{SEq:EOM_Del_Class_HInt_Vector} as
\begin{equation}
\Delta(t) = \Delta(0) e^{-2  g J^{z}_0 t}.
\label{SEq:EOM_Del_Class_HInt_Vector_Sol} 
\end{equation} 
We now define a rotating frame by  
\begin{equation}
    \widetilde{\bm{S}}_j (t)     =     \mathbb{R}_{\bm{z}} (  - 2  g J^{z}_0 t   ) \cdot \bm{S}_j (t).
\label{SEq:Rotating_Frame_HInt} 
\end{equation} 
In this frame, we have 
\begin{equation}
    \begin{split}
    \widetilde{\Delta} (t) &=  g  \sum_{k=1}^{N}  \widetilde{S}_{j}^{-} = g  e^{ + 2  g J^{z}_0 t}  \sum_{k=1}^{N} S_{j}^{-}  \\
    &=   e^{ + 2  g J^{z}_0 t} \Delta(0) e^{-2  g J^{z}_0 t} = \Delta(0) = \widetilde{\Delta} (0).
    \end{split}
\label{SEq:Rotating_Frame_HInt_Delta} 
\end{equation}
This shows that the order parameter in the rotating frame $\widetilde{\Delta}$ remains constant. Substituting Eq.\ \eqref{SEq:Rotating_Frame_HInt} in the Hamilton equations in Eq.\ \eqref{SEq:Tan_Map_HInt}, we obtain the new equations of motion in the rotating as
\begin{equation}
    \frac{    d   \widetilde{\bm{S}}_j     }{    dt    }    =    -2 \underbrace{    \left(     \bm{\Delta}(0)     +    g J^{z}_0 \bm{z}   \right)    }_{    \equiv \bm{\Delta}_\mathrm{rot}    }    \times    \widetilde{\bm{S}}_j.
\label{SEq:EOM_Class_HInt_Rot}
\end{equation}
From Eq.\ \eqref{SEq:Tan_Map_HInt}, we derive that \(\Delta(t) = \Delta(0) e^{-2  g J^{z}_0 t}\). As a result, \(\bm{\Delta}\) is not a constant of motion for \(H_\mathrm{int}\). However, since the definition of $\bm{\Delta}_\mathrm{rot}$ contains only the initial vector \(\bm{\Delta}(0)\), it is a time-independent constant. As a result, above is a simple Bloch equation, whose solution can be readily obtained as 
\begin{equation}
    \widetilde{\bm{S}}_j^\prime = \mathbb{R}_{\bm{n}_\mathrm{rot}} (  - 2  \lvert \bm{\Delta}_\mathrm{rot} \rvert t   ) \cdot \widetilde{\bm{S}}_j, \; \textrm{with} \;\; \bm{n}_\mathrm{rot} = \frac{\bm{\Delta}_\mathrm{rot}}{\lvert \bm{\Delta}_\mathrm{rot} \rvert}. 
\label{SEq:EOM_Class_HInt_Rot_Soln}
\end{equation}
From Eq.\ \eqref{SEq:Rotating_Frame_HInt}, we observe that $\widetilde{\bm{S}}_j(0) = \bm{S}_j(0)$. Using Eqs. \eqref{SEq:Rotating_Frame_HInt} and \eqref{SEq:EOM_Class_HInt_Rot_Soln}, we write the solution to Eq.\ \eqref{SEq:Tan_Map_HInt} as
\begin{equation}
e^{L_\mathrm{int}\tau}\colon \begin{cases}
    \bm{S}_j^\prime    =     \mathbb{R}_{\bm{z}} (  2  g J^{z}_0 \tau   ) \cdot \mathbb{R}_{\bm{n}_\mathrm{rot}} (  - 2  \lvert \bm{\Delta}_\mathrm{rot} \rvert \tau   ) \cdot \bm{S}_j, \\[5pt]
    \begin{split}
    \delta\bm{S}_j^\prime = \bigg[ \delta\mathbb{R}_{\bm{z}} (  2  g J^{z}_0 \tau   ) \bigg] \cdot \mathbb{R}_{\bm{n}_\mathrm{rot}} (  - 2  \lvert \bm{\Delta}_\mathrm{rot} \rvert \tau   ) \cdot \bm{S}_j \\
    + \mathbb{R}_{\bm{z}} (  2  g J^{z}_0 \tau   ) \cdot \bigg[ \delta\mathbb{R}_{\bm{n}_\mathrm{rot}} (  - 2  \lvert \bm{\Delta}_\mathrm{rot} \rvert \tau   ) \bigg] \cdot \bm{S}_j \\
    \mathbb{R}_{\bm{z}} (  2  g J^{z}_0 \tau   ) \cdot \mathbb{R}_{\bm{n}_\mathrm{rot}} (  - 2  \lvert \bm{\Delta}_\mathrm{rot} \rvert \tau   ) \cdot \delta\bm{S}_j,
    \end{split}
\end{cases}
\label{SEq:Tan_Map_Soln_HInt}
\end{equation}
where, instead of directly integrating, we obtain the solution to the variation equation of \eqref{SEq:Tan_Map_HInt} by differentiating the solution to the Hamilton equations. 

\subsubsection{Expressions for $\delta\mathbb{R}$ in Eq.\ \eqref{SEq:Tan_Map_Soln_HInt} }
\label{Sec:Further_Simpl_in_Interacting_Part_BCS_Soln}

We further simplify the solution to the deviation equations in \eqref{SEq:Tan_Map_Soln_HInt} by obtaining explicit expressions for $\delta\mathbb{R}_{\bm{z}} (  2  g J^{z}_0 \tau   )$ and $\delta\mathbb{R}_{\bm{n}_\mathrm{rot}} (  - 2  \lvert \bm{\Delta}_\mathrm{rot} \rvert \tau   )$ in Eqs.\ \eqref{SEq:Deriv_Del_CW_Rot_Mat} and \eqref{SEq:Deriv_z_CCW_Rot_Mat}, respectively. To that end, we use the following angle-axis representation of the rotation matrices:
\begin{subequations}
\begin{align}
    \mathbb{R}_{\bm{n}} (  \theta ) &= \mathbb{1} + \sin\theta\; \mathbb{K}(\bm{n}) + (1 - \cos\theta)\;\mathbb{K}^2(\bm{n}), \\
    \mathbb{K}(\bm{n}) &= \begin{pmatrix} 0 & -n_z & n_y \\ n_z & 0 & -n_x  \\ -n_y & n_x & 0 \end{pmatrix},
\end{align}
\label{SEq:Angle-Axis} 
\end{subequations}
where $n_\alpha$'s for $\alpha = x, y,$ and $z$ denote different components of the unit vector $\bm{n}$ corresponding to axis of rotation. For the ease of future calculations, we define:
For the ease of future calculations, we define:
\begin{equation}
    \theta_{\bm{n}_\mathrm{rot}} = - 2 \lvert \bm{\Delta}_\mathrm{rot} \rvert \tau, \qquad \theta_{\bm{z}} = 2  g J^{z}_0 \tau. 
\label{SEq:Defn_Rot_Angle}
\end{equation}   
This obtains  
\begin{equation}
    \delta\theta_{\bm{n}_\mathrm{rot}} = - 2\tau \; \delta\lvert \bm{\Delta}_\mathrm{rot} \rvert, \qquad \delta \theta_{\bm{z}} = 2g\tau \; \delta J^{z}_0, 
\label{SEq:Dev_Rot_Angle}
\end{equation}
where we used the following: 
\begin{equation}
    \begin{split}
    \delta \lvert \bm{\Delta}_\mathrm{rot} \rvert = \frac{  \Delta_x \delta \Delta_x + \Delta_y \delta \Delta_y + g^2 J^{z}_0 \delta J^{z}_0  }{  \lvert \bm{\Delta}_\mathrm{rot} \rvert  }.
    \end{split}
\label{SEq:Dev_Rot_Angle_Auxiliary}
\end{equation}
Identifying 
\begin{equation}
    \begin{split}
    \bm{n}_\mathrm{rot} &= \frac{  \Delta_x  }{  \lvert \bm{\Delta}_\mathrm{rot} \rvert  }  \bm{x}  +  \frac{  \Delta_y  }{  \lvert \bm{\Delta}_\mathrm{rot} \rvert  }  \bm{y}  +  \frac{  g J^{z}_0  }{  \lvert \bm{\Delta}_\mathrm{rot} \rvert  }  \bm{z} \\   
    &\equiv    n_\mathrm{rot, x}  \bm{x}  +  n_\mathrm{rot, y}  \bm{y}  +  n_\mathrm{rot, z}  \bm{z}, 
    \end{split}
\label{SEq:Defn_Unit_Vec_Comps}    
\end{equation} 
we obtain 
\begin{subequations}
\begin{align}
    \delta n_\mathrm{rot, x}     &=     \frac{ \delta \Delta_x }{ \lvert \bm{\Delta}_\mathrm{rot} \rvert } - \frac{ \Delta_x }{ \lvert \bm{\Delta}_\mathrm{rot} \rvert^2 }\delta\lvert \bm{\Delta}_\mathrm{rot} \rvert, \\
    \delta n_\mathrm{rot, y}     &=     \frac{ \delta \Delta_y }{ \lvert \bm{\Delta}_\mathrm{rot} \rvert } - \frac{ \Delta_y }{ \lvert \bm{\Delta}_\mathrm{rot} \rvert^2 }\delta\lvert \bm{\Delta}_\mathrm{rot} \rvert, \\
    \delta n_\mathrm{rot, z}     &=     g  \left(  \frac{ \delta J^{z}_0 }{ \lvert \bm{\Delta}_\mathrm{rot} \rvert } - \frac{ J^{z}_0 }{ \lvert \bm{\Delta}_\mathrm{rot} \rvert^2 }\delta\lvert \bm{\Delta}_\mathrm{rot} \rvert  \right).
\end{align}
\label{SEq:Deriv_Unit_Vec_Comps}
\end{subequations}
Finally, using Eqs.\ \eqref{SEq:Angle-Axis}, \eqref{SEq:Defn_Rot_Angle}, \eqref{SEq:Dev_Rot_Angle}, \eqref{SEq:Defn_Unit_Vec_Comps}, and \eqref{SEq:Deriv_Unit_Vec_Comps}, we obtain
\begin{multline}
\delta \mathbb{R}_{\bm{n}_\mathrm{rot}} (  \theta_{\bm{n}_\mathrm{rot}}   ) = \delta\theta_{\bm{n}_\mathrm{rot}}\cos\theta_{\bm{n}_\mathrm{rot}} \mathbb{K}(\bm{n}_\mathrm{rot}) \\     +    \sin\theta_{\bm{n}_\mathrm{rot}} \delta \mathbb{K}(\bm{n}_\mathrm{rot})    +     \delta\theta_{\bm{n}_\mathrm{rot}}\sin\theta_{\bm{n}_\mathrm{rot}} \mathbb{K}^2(\bm{n}_\mathrm{rot}) \\
+ \left(1 - \cos\theta_{\bm{n}_\mathrm{rot}} \right) \bigg[ \mathbb{K}(\bm{n}_\mathrm{rot}) \cdot \delta \mathbb{K}(\bm{n}_\mathrm{rot})   \\
+   \delta \mathbb{K}(\bm{n}_\mathrm{rot}) \cdot \mathbb{K}(\bm{n}_\mathrm{rot})    \bigg],    
\label{SEq:Deriv_Del_CW_Rot_Mat}
\end{multline}
where
\begin{equation}
\begin{split}
    \mathbb{K}(\bm{n}_\mathrm{rot}) &= \begin{pmatrix} 0 & -n_\mathrm{rot, z} & n_\mathrm{rot, y} \\ n_\mathrm{rot, z} & 0 & -n_\mathrm{rot, x}  \\ -n_\mathrm{rot, y} & n_\mathrm{rot, x} & 0 \end{pmatrix},      \\[10pt]      \delta\mathbb{K}(\bm{n}_\mathrm{rot}) &= \begin{pmatrix} 0 & -\delta n_\mathrm{rot, z} & \delta n_\mathrm{rot, y} \\ \delta n_\mathrm{rot, z} & 0 & -\delta n_\mathrm{rot, x}  \\ -\delta n_\mathrm{rot, y} & \delta n_\mathrm{rot, x} & 0 \end{pmatrix}.
\end{split}
\label{SEq:Deriv_K_Mat}
\end{equation}
Similarly, using Eqs.\ \eqref{SEq:Angle-Axis}, \eqref{SEq:Defn_Rot_Angle}, and \eqref{SEq:Dev_Rot_Angle}, we obtain 
\begin{equation}
    \delta \mathbb{R}_{\bm{z}} (  \theta_{\bm{z}}   ) = \delta\theta_{\bm{z}}\cos\theta_{\bm{z}} \mathbb{K}(\bm{z})  +   \delta\theta_{\bm{z}}\sin\theta_{\bm{z}} \mathbb{K}^2(\bm{z}),
\label{SEq:Deriv_z_CCW_Rot_Mat} 
\end{equation}
where 
\begin{equation}
\mathbb{K}(\bm{z}) = \begin{pmatrix} 0 & -1 & 0 \\ +1 & 0 & 0  \\ 0 & 0 & 0 \end{pmatrix}.
\end{equation}

\subsection{Symplectic Integration Scheme: $\mathrm{SABA}_2$}
\label{Sec:SABA2_Intro}

The symplectic integrator used in all our numerics is called $\mathrm{SABA}_2$. It is described by the following product of resolvent operators $e^{L_\mathrm{free}\tau}$ and $e^{L_\mathrm{int}\tau}$ of addends $H_\mathrm{free}$ and $H_\mathrm{int}$, respectively:
\begin{equation}
\begin{split}
  \mathrm{SABA}_2(\tau ) &= e^{a_1 \tau L_\mathrm{free}}e^{b_1 \tau L_\mathrm{int}}e^{a_2 \tau L_\mathrm{free}} e^{b_1 \tau L_\mathrm{int}}e^{a_1 \tau L_\mathrm{free}} 
\end{split}
\label{SEq:saba2}
\end{equation}
for a given time-step $\tau$. Expressions for $e^{L_\mathrm{free}\tau}$ and $e^{L_\mathrm{int}\tau}$ are provided in Eqs.\ \eqref{SEq:Tan_Map_Soln_HFree} and \eqref{SEq:Tan_Map_Soln_HInt}, respectively.   
The coefficients $\{a_1,a_2,b_1\}$ are
\begin{equation}
\begin{split}
&a_1 = \frac{1}{2}\left(1-\frac{1}{\sqrt{3}} \right), 
 \qquad \ \ 
a_2=\frac{1}{\sqrt{3}},
 \qquad \ \ 
b_1 = \frac{1}{2}. \\ 
\end{split}
\label{SEq:saba2_coeff}
\end{equation}
Using the Baker–Campbell–Hausdorff formula, it can be proven \cite{Laskar_sup} that 
\begin{equation}
\begin{split}
  \mathrm{SABA}_2(\tau ) &= e^{\tau L_\mathrm{BCS}} + \mathcal{O}(\tau^3), \\
  K_\mathrm{BCS} &= H_\mathrm{BCS} + \mathcal{O}(\tau^2),  
\end{split}
\label{SEq:saba2_Err1}
\end{equation}
where \(K_\mathrm{BCS}\) is the effective Hamiltonian corresponding to the time evolution operator \( \mathrm{SABA}_2(\tau ) \), and \(L_\mathrm{BCS}\) is the Liouvillian operator for the original Hamiltonian \(H_\mathrm{BCS}\).  

\subsection{Particle-Hole Symmetry of $H_\mathrm{BCS}$}
\label{Sec:Particle-Hole_Symmetry}

The BCS Hamiltonian and its trotterized dynamics obey the particle-hole symmetry. Here we explicitly describe that symmetry as 
\begin{equation}
    \begin{split}
    S^x(\varepsilon) &= S^x(-\varepsilon), \\ S^y(\varepsilon) &= - S^y(-\varepsilon), \\ S^z(\varepsilon) &= - S^z(-\varepsilon).
    \end{split}
\label{SEq:PH_Def1}
\end{equation}
Given that the $N$ single-particle energies $\varepsilon_j$ are symmetrically distributed within the interval $[-W/2, +W/2]$, the symmetry for even $N$ can be expressed as follows:  
\begin{equation}
    \begin{split}
    S^x_j &= S^x_{N-j+1}, \\ 
    S^y_j &= - S^y_{N-j+1}, \\ 
    S^z_j &= - S^z_{N-j+1}.
    \end{split}
\label{SEq:PH_Def2}
\end{equation}
The BCS spin equations, along with the spin evolution equations derived from both the free and interacting components of the BCS Hamiltonian, also respect this symmetry.

\section{Extrema of the BCS Energy Band}
\label{Sec:Extrema_BCS_E_Band}

\begin{figure}
    \centering
    \includegraphics[trim={1.00cm 0.75cm 0.0cm 0.0cm},clip,scale=0.70]{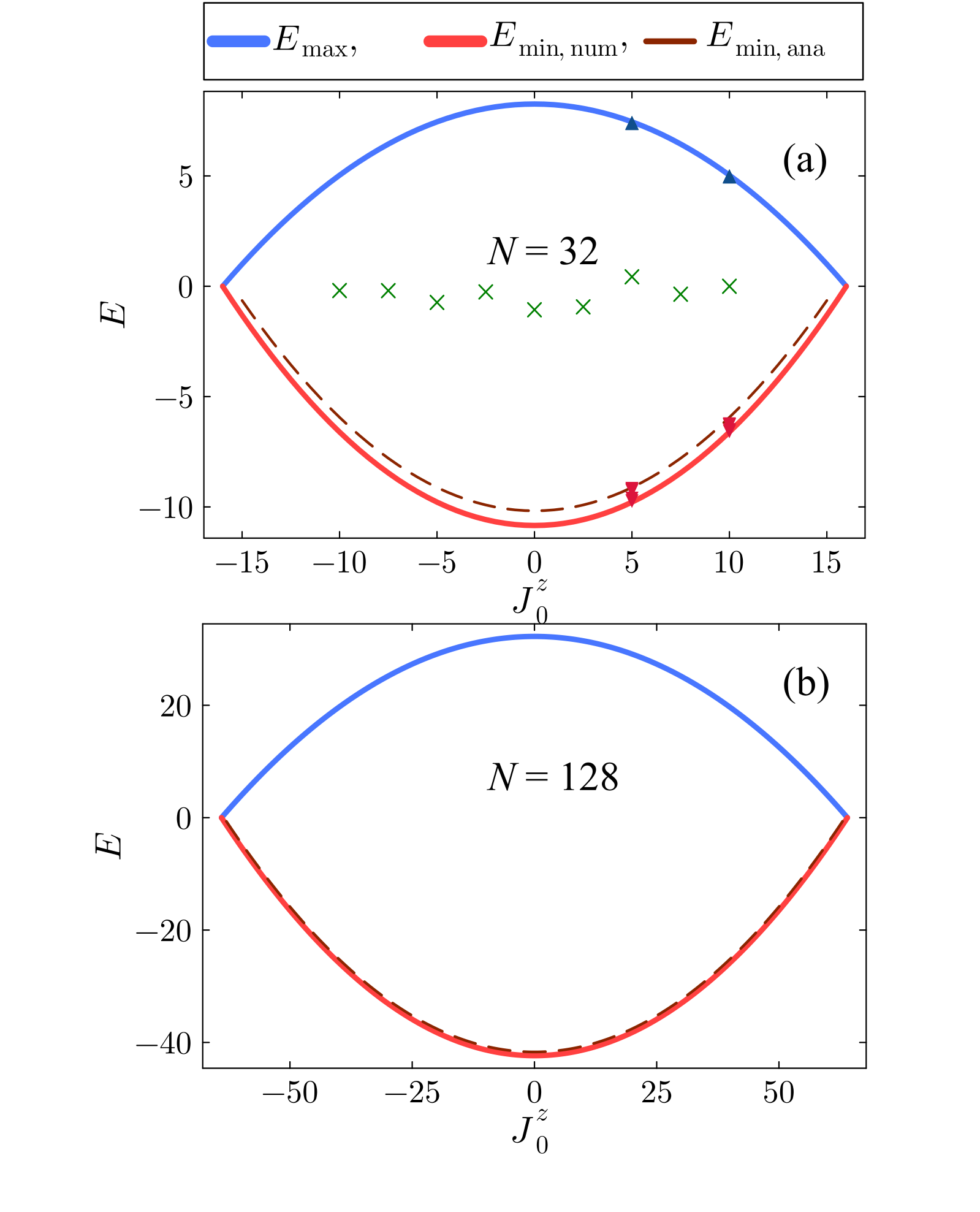}
    \caption{ We show the energy extrema for $N = 32$ and $N = 128$ with various $J_0^{z}$ in (a) and (b), respectively. The energy maxima $E_\mathrm{max}$ corresponding to Eq.\ \eqref{SEq:Maximum_Energy_BCS} are shown with the solid blue line. The solid red line shows the numerically obtained energy minima $E_\mathrm{min, num}$ for different $J_0^z$. The energy minima $E_\mathrm{min, ana}$ obtained from the analytical expression in Eq.\ \eqref{SEq:EMin_Ana_BCS}, obtained by taking the $N \to \infty$ limit, is shown by the brown dashed line. Indeed, the analytical results converge more closely to the numerical ones as we increase $N$. In panel (a), we denote the energies of the spin configurations used in Figs.\ (1), (2), and (3) of the main text by green crosses. The blue up triangles and the red down triangles show the energies of the perturbed extrema configurations that are considered in Fig.\ \ref{SFig:No_Prethermalization_Large_tau}.  }
    \label{SFig:BCS_E_Ana_Num_Increasing_N_Comp}
\end{figure}

Recalling that the BCS Hamiltonian,  
\begin{equation}
    \label{SEq:H_BCS_1}
    \begin{split}
    H_\mathrm{BCS} &= \underbrace{    \sum_{j=1}^{N} 2\varepsilon_j S_j^{z}    }_{ H_\mathrm{free} }     \underbrace{    - g \sum_{j,k = 1}^{N} S_{j}^{+} S_{k}^{-}    }_{ H_\mathrm{int} } \\
    &= \sum_{j=1}^{N} 2\varepsilon_j S_j^{z} \underbrace{ - g \left[ \left(J^{x}\right)^{2} + \left(J^{y}\right)^{2} \right] }_{ H_\mathrm{int} = - g|J^\perp|^2 }, 
    \end{split}
\end{equation}
conserves the total $z$-component of the spins $J^{z}$, we aim to determine the minimum and maximum possible energy values for a specific initial $J^{z}$ value,  denoted as $J^{z}_0$. To that end, writing the spins using the polar coordinate as
\begin{equation}
    \begin{split}
    S_j^{x} &= S \sin\theta_i\cos\phi_i, \\ 
    S_j^{y} &= S \sin\theta_i\sin\phi_i, \\ 
    S_j^{z} &= S \cos\theta_i, 
    \end{split}
    \label{SEq:Polar}
\end{equation} 
with $S = \lvert \bm{S}_j \rvert = 1/2$, we need to optimize $H$ -- i.e., find either maximum or minimum -- subject to the constraint $J^{z} = J^{z}_{0}$, where $J^{z}_{0}$ is a constant. 

We note that rotating the $j$th spin around the $z$-axis does not change $H_\mathrm{free}$, while the second term is just $- g|J^\perp|^2$. Thus, to obtain the minimum / maximum energy, we only need $J^\perp$ to be maximal / minimal. The $J^\perp$ is maximum when the projections of all spins onto $xy$-plane point along the same direction. Without loss of generality, we can take this direction as the positive $x$-axis. On the other hand, $J^\perp$ attains its minimum $J^\perp = 0$, when all the spins point in the $z$-direction -- see below. Therefore, all the spins in these extrema configurations are co-planar. For these extrema spin configurations, using $\phi_j = 0$, we simplify the polar representations of the spins \eqref{SEq:Polar} as 
\begin{subequations}
\begin{gather}
    S_j^{x} = \frac{1}{2} \sin\theta_i, \; S_j^{y} = 0, \; S_j^{z} = \frac{1}{2} \cos\theta_i, \label{SEq:Polar_Simplified}\\
    J^{\perp} = \sum_{j = 1}^{N} S_j^{x}, \;\; J^{z} = \sum_{j = 1}^{N} S_j^{z}.
    \label{SEq:J_Comp_Defn}
\end{gather}
\label{SEq:Polar_Simplified_J_Comp_Defn}
\end{subequations}
    
To find the minimum / maximum energy, we need to solve for $\theta_i$ from
\begin{equation}
\frac{\partial H}{\partial \theta_j} = 2\mu \frac{\partial J^{z}}{\partial \theta_j}, 
\label{SEq:Lagrange_Multiplier_Defn}
\end{equation}
where $2\mu$ is the Lagrange multiplier corresponding to the angular momentum constraint. Using Eqs.\ \eqref{SEq:H_BCS_1}, \eqref{SEq:Polar_Simplified}, and \eqref{SEq:Lagrange_Multiplier_Defn}, we obtain
\begin{subequations}
\begin{equation}
    (\mu - \varepsilon_j) \sin\theta_j = gJ^\perp \cos\theta_j, 
\end{equation}
\begin{equation}
    \implies 
    \begin{cases}
    \sin\theta_j = \frac{   g J^\perp   }{   \sqrt{   (\mu - \varepsilon_j)^2 + \left(g J^{\perp}\right)^2    }   }, \\[10pt] 
    \cos\theta_j = \frac{   (\mu - \varepsilon_j)      }{   \sqrt{   (\mu - \varepsilon_i)^2 + \left(g J^{\perp}\right)^2    }   }.
    \end{cases}
    \label{SEq:Extremizing_Soln_theta_Fin}
\end{equation}
\label{SEq:Extremizing_Soln_theta}
\end{subequations}

Recall our argument above Eq.\ \eqref{SEq:Polar_Simplified_J_Comp_Defn}, the maximum energy state is characterized by $J^\perp = 0$. Using this in Eq.\ \eqref{SEq:Extremizing_Soln_theta_Fin}, implies that in the maximum energy state $\theta_j = 0$. In particular, the maximum energy state for a general $J^z_0$ is the state where the first $K$ (lowest energy) spins are down, and the remaining $N-K$ spins are up, where $K$ is determined by the equation $-K/2 + (N-K)/2 = J^z_0$, i.e., $K = N/2 - J^z_0$ 
\begin{equation}
    -K/2 + (N-K)/2 = J^z_0 \implies K = N/2 - J^z_0.  
    \label{SEq:Maximum_Energy_Spin_Config_K}
\end{equation}
Therefore, the expression energy maximum for a general $J^z_0$ is given by
\begin{equation}
    \begin{split}
    E_\mathrm{max} &= -\left( \varepsilon_1 + \cdots + \varepsilon_K  \right) + \left( \varepsilon_{K+1} + \cdots + \varepsilon_N  \right) \\
    &= \frac{N^2 - 4\left(J_0^z\right)^2}{4(N-1)},
    \end{split}
\label{SEq:Maximum_Energy_BCS}
\end{equation}
where the value of $K$ is given in Eq.\ \eqref{SEq:Maximum_Energy_Spin_Config_K}. Also, since we have chosen equidistant single-particle energies $\varepsilon_j$ in the range $[-1/2, +1/2]$ with single-particle level spacing $\delta = 1/(N-1)$, we have 
\begin{equation}
    \varepsilon_{j} = -\frac{1}{2} + \frac{j-1}{N-1}.  
    \label{SEq:Single-Partcle_Energy}
\end{equation}

In the minimum-energy configuration, we have $J^\perp > 0$. Using Eqs.\ \eqref{SEq:J_Comp_Defn} and \eqref{SEq:Extremizing_Soln_theta_Fin} in Eq.\ \eqref{SEq:J_Comp_Defn}, we obtain the following self-consistent equations: 
\begin{subequations} 
\begin{align}
    \sum_{i = 1}^{N} \frac{   (\mu - \varepsilon_i)      }{   2\sqrt{ (\mu - \varepsilon_i)^2 + \left(gJ^{\perp}\right)^{2}   }   } &= J^{z}_0, \label{SEq:Chem_Pot_Eq} \\
    \sum_{i = 1}^{N} \frac{   g J^{\perp}   }{   2\sqrt{ (\mu - \varepsilon_i)^2 + \left(gJ^{\perp}\right)^{2}   }   } &= J^{\perp}. \label{SEq:Gap_Eq}
\end{align}
\label{SEq:Self_Consistent_Equations}
\end{subequations}
The above \eqref{SEq:Self_Consistent_Equations} are two coupled transcendental equations for $\mu$ and $\Delta = gJ^{\perp}$. Among these two equations, Eq.\ \eqref{SEq:Gap_Eq} becomes the superconducting gap equation when $J^{z}_0 = 0$. In that case, $\mu$ must also be zero, which satisfies Eq.\ \eqref{SEq:Chem_Pot_Eq}.

To obtain $E_\mathrm{min, num}$, we solve for $\mu$ and $J^\perp$ from Eqs.\ \eqref{SEq:Chem_Pot_Eq} and \eqref{SEq:Chem_Pot_Eq}. We then substitute these values in Eqs.\ \eqref{SEq:Extremizing_Soln_theta_Fin} and \eqref{SEq:Polar_Simplified} to obtain all the spin vectors. Finally, the quantity $E_\mathrm{min, num}$ represents the value of $H_\mathrm{BCS}$ calculated from Eq.\ \eqref{SEq:H_BCS_1} calculated from Eq.\ \eqref{SEq:H_BCS_1} for the aforementioned spin configuration.

To obtain an analytical expression of the energy minimum, we replace the summation over the spin indexes by an integration as
\begin{equation}
    \sum_{j = 1}^{N} \longrightarrow \frac{1}{\delta}\int_{-1/2}^{1/2} d\varepsilon = (N-1)\int_{-1/2}^{1/2} d\varepsilon. 
    \label{SEq:Sum_to_Int_E_BCS}
\end{equation}
By making the above transformation in Eq.\ \eqref{SEq:Self_Consistent_Equations}, we obtain   
\begin{subequations}
\begin{align}
    \mu &= \frac{J^{z}_0}{N-1}  \left( \frac{e^2 + 1}{e^2 - 1} \right), \label{SEq:mu_BCS_Self_Consistent_Soln} \\
    \Delta &= gJ^{\perp} = \frac{ e \sqrt{  (e^2 + 1)^2 - 4(e^2 - 1)^2\mu^2  } }{e^4 - 1}, \label{SEq:Delta_BCS_Self_Consistent_Soln}
\end{align}
\label{SEq:Self_Consistent_Equations_Soln}
\end{subequations}
where $e \approx 2.7183$ is the Euler's number. Using the substitution \eqref{SEq:Sum_to_Int_E_BCS}, from Eq.\ \eqref{SEq:H_BCS_1}, we obtain the following:
\begin{widetext}
\begin{multline}
    E_\mathrm{min, ana} (N, J^{z}_0) = -\frac{(N-1)}{8} \Big[ (1+2\mu) \sqrt{4\Delta^2 + (1-2\mu)^2} + (1-2\mu) \sqrt{4\Delta^2 + (1+2\mu)^2} \\ + 2\Delta^2 \Big\{ \ln\Big[\left( 2\mu -1 + \sqrt{ 4\Delta^2 + (1-2\mu)^2 } \right) \left(-1 -2\mu + \sqrt{ 4\Delta^2 + (1+2\mu)^2 }  \right)\Big] \\
- \ln\Big[\left( 1-2\mu + \sqrt{ 4\Delta^2 + (1-2\mu)^2 } \right) \left(1 + 2\mu + \sqrt{ 4\Delta^2 + (1+2\mu)^2 }  \right)\Big] \Big\} \Big]- \underbrace{\Delta^2 (N-1),}_{g \sum_{j,k = 1}^{N} S_{j}^{+} S_{k}^{-}}
\label{SEq:EMin_Ana_BCS}
\end{multline}
\end{widetext}
where $\mu$ and $\Delta$ are given in Eqs.\ \eqref{SEq:mu_BCS_Self_Consistent_Soln} and \eqref{SEq:Delta_BCS_Self_Consistent_Soln}, respectively.

\section{\(J_0^z\) Dependence of the Lyapunov Spectrum}
\label{Sec:Jz0_Dep_LS}

\begin{figure}
    \centering
    \includegraphics[trim={0.0cm 0.0cm 0cm 0cm},clip,scale=0.60]{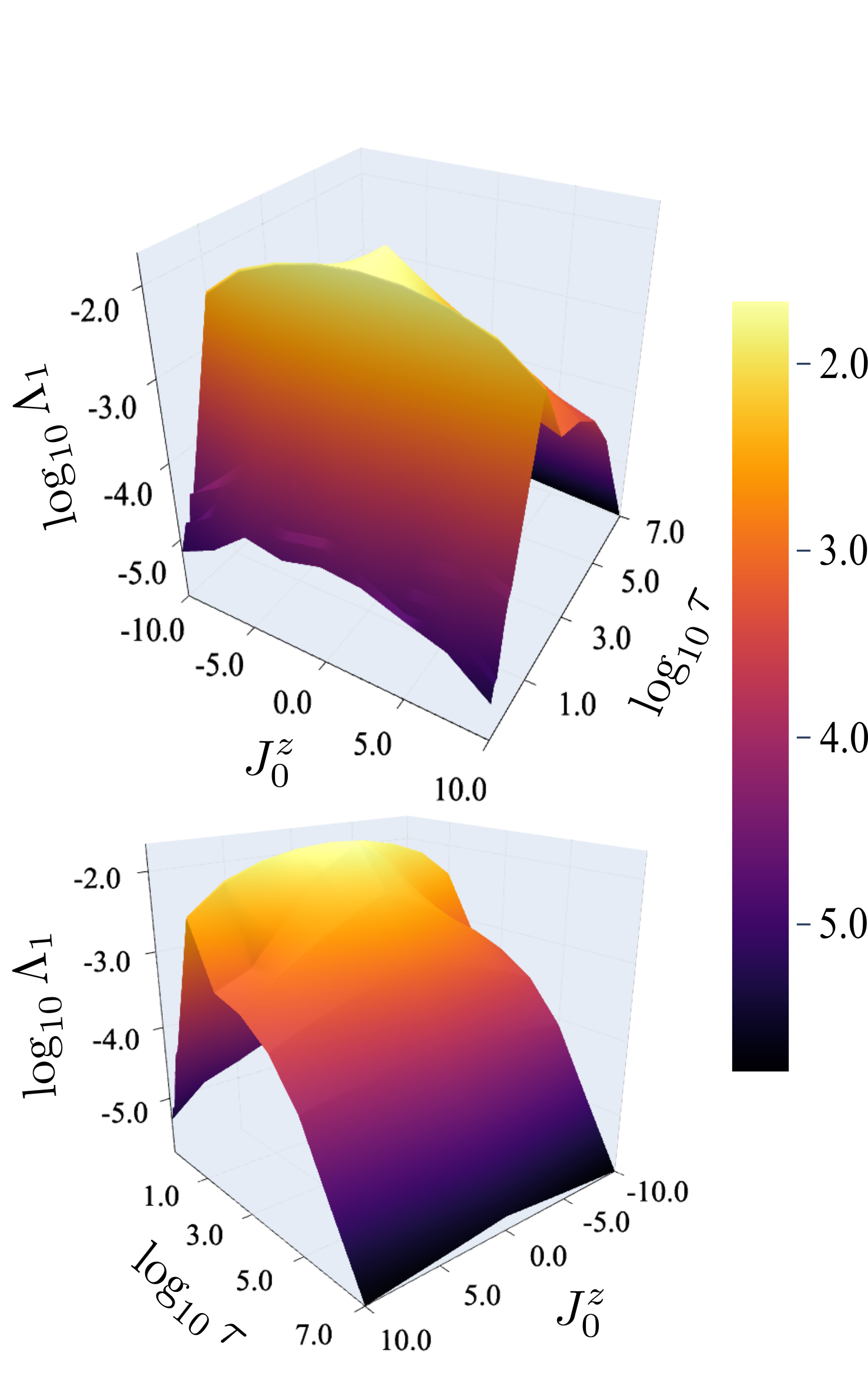}
    \caption{ In this surface plot, we show $\log_{10} \Lambda_1$ as a function of $\log_{10} \tau$ and $J^{z}_0$ for $N = 32$. We show the data for $J_0^z = -10, -7.5, -5, \ldots, +10$. The top and bottom figures show two different orientations of the same surface plot. The color bar indicates the values of $\log_{10} \Lambda_1$. The $\log_{10} \Lambda_1$ vs $\log_{10} \tau$ curve for $J_0^z = 0$, i.e., the cross section of this surface plot at $J_0^z = 0$, is quite similar to the blue curve shown in Fig.\ 1 of the main text. }
    \label{SFig:3D_MLE}
\end{figure} 

\begin{figure}
    \centering
    \includegraphics[trim={0.75cm 0.0cm 0cm 0cm},clip,scale=0.85]{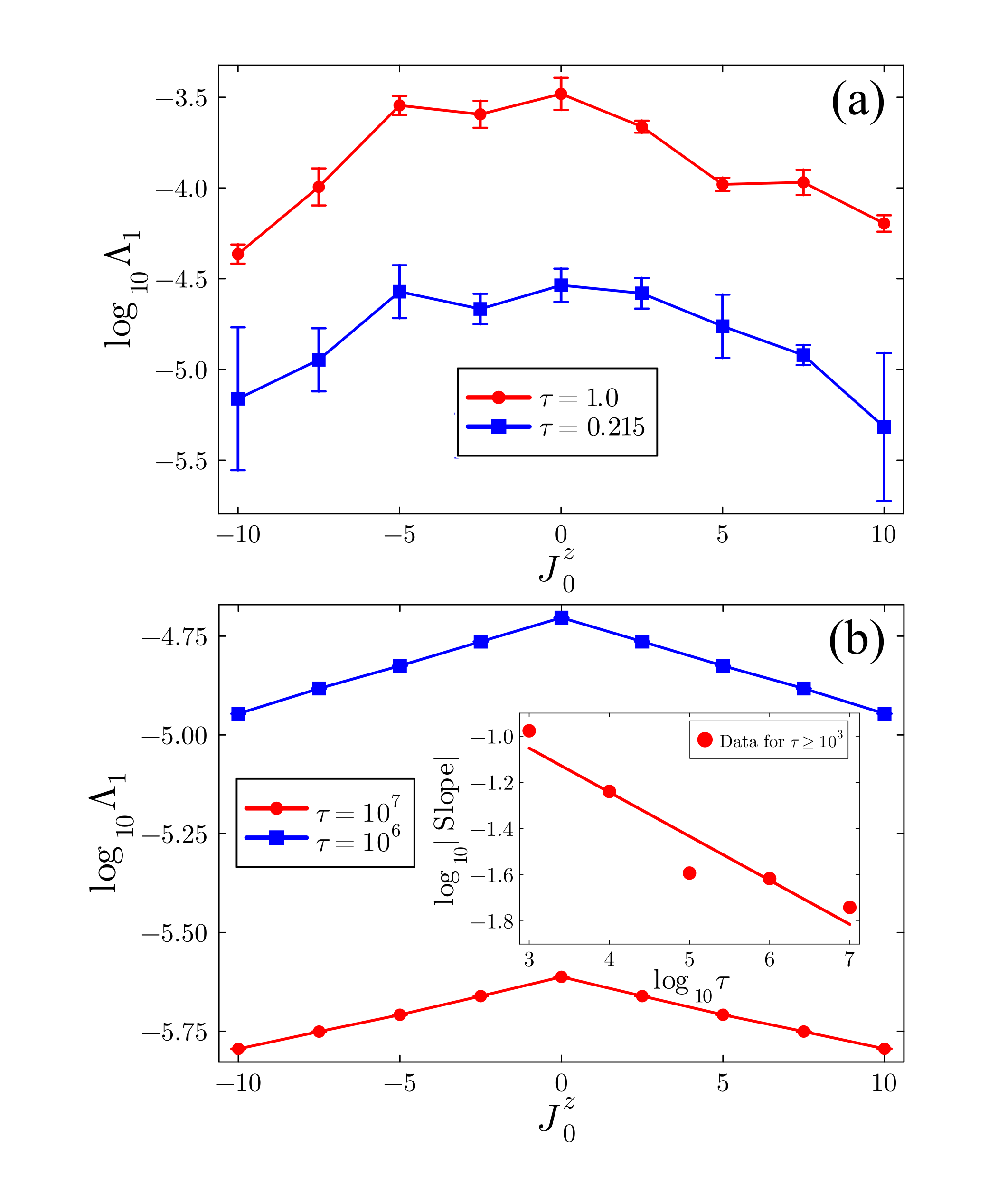}
    \caption{ We show the dependence of $\log_{10} \Lambda_1$ on $J^{z}_0$ by extracting the cross-sectional views of the surface plot in Fig.\ \ref{SFig:3D_MLE} at fixed $\tau$ values. Panel (a) and (b) show data in the weakly chaotic and in the memoryless regimes, respectively. The blue circles and the red squares in panel (a) correspond to step sizes $\tau = 0.215$ and $\tau  = 1.0$, whereas the blue circles and the red squares in panel (b) correspond to step sizes $\tau = 10^6$ and $\tau  = 10^7$. The curves in panel (a) show slight asymmetry around $J_0^z = 0$, whereas the curves in panel (b) display perfect reflection symmetry about $J_0^z = 0$. We obtain the slopes ($<0$) by fitting the ($\log_{10} \Lambda_1$, $\vert J^{z}_0 \vert$) data for $\tau > 10^3$ to a linear model: $y = mx + c$. In the inset of panel (b), we show the $\log$-$\log$ plot of $\vert m \vert$ versus $\tau$, where the linear fit is described by $y = -0.19 x - 0.48$. }
    \label{SFig:MLE_Reflection}
\end{figure} 

\begin{figure*}
    \centering
    \includegraphics[trim={0.75cm 0.5cm 0cm 0cm},clip,scale=0.90]{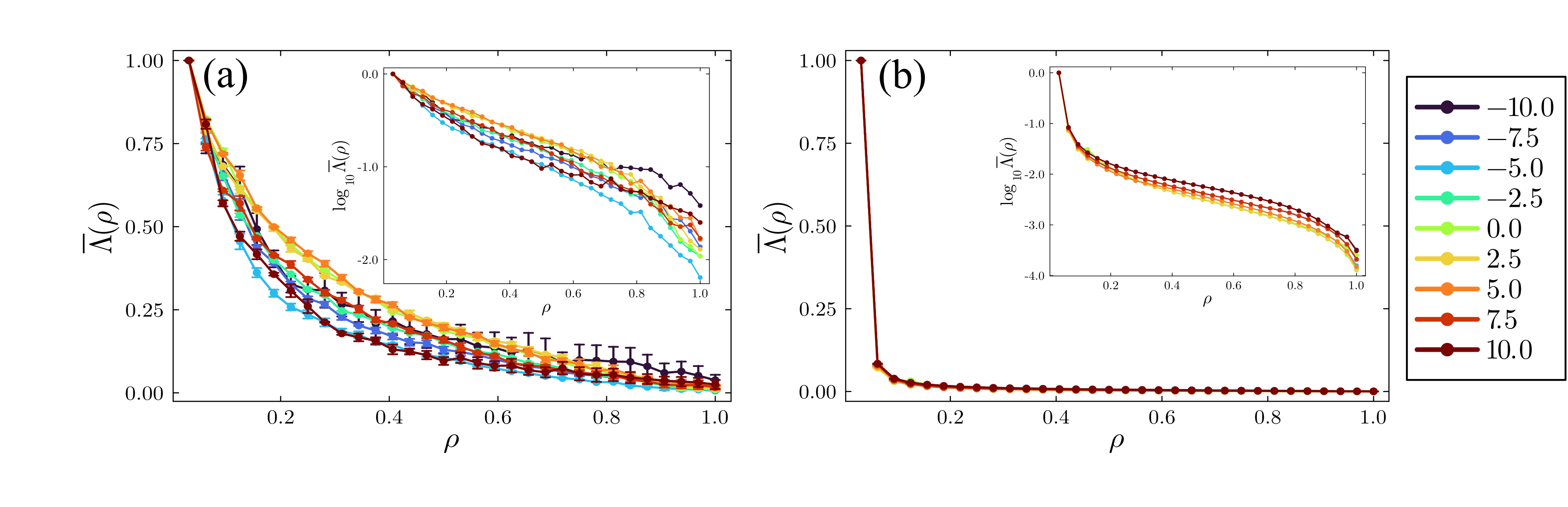}
    \caption{ We show rescaled Lyapunov spectra for several different $J^{z}_0$ for $N = 32$ for $\tau = 1.0$ and $\tau = 10^7$ in panels (a) and (b) respectively. The plots display the rescaled Lyapunov exponents, $\overbar{\Lambda}(\rho) \equiv \Lambda_i/\Lambda_1$, on the $y$-axis for $i = 1, \ldots, N$, against the normalized index $\rho \equiv i/N$ on the $x$-axis. In these plots, different colors represent different initial spin configurations, each characterized by a specific $J_0^z$ value -- cf. the caption of Fig.\ \ref{SFig:3D_MLE}. The corresponding $J_0^z$ values are listed in the legend on the right. In the insets, we plot $\log_{10} \overbar{\Lambda}(\rho)$ against $\rho$. None of these data sets in the two insets conform to a linear model.  }
    \label{SFig:Rescaled_LS_vs_Jz}
\end{figure*}

\begin{figure*}
    \centering
    \includegraphics[trim={0.75cm 0.75cm 0cm 0cm},clip,scale=0.85]{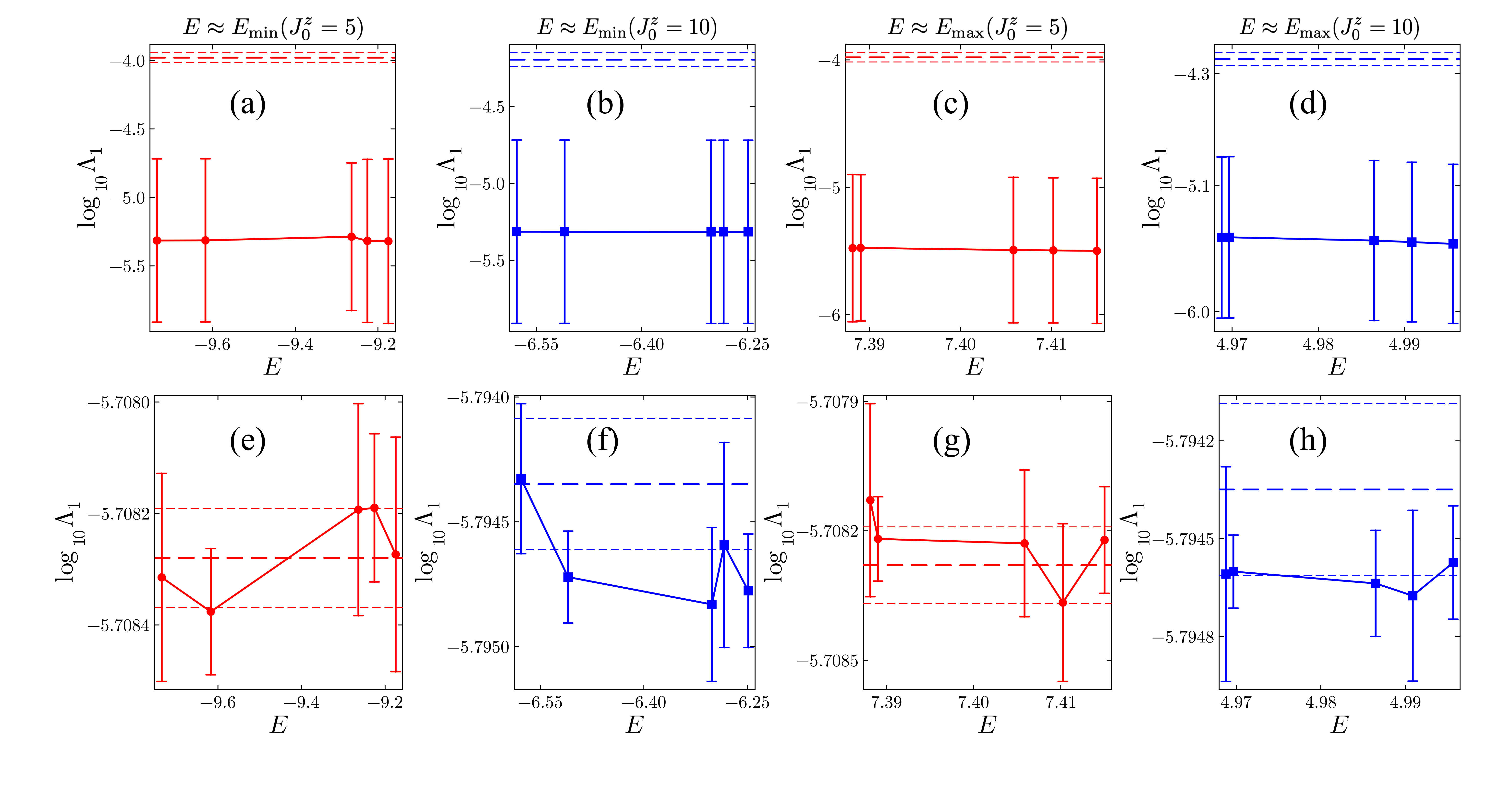}
    \caption{ Variation of $\log_{10} \Lambda_1$ with $E$ close to the edges of the $E(J_0^z)$ band for $\tau = 1$ (first row) and $\tau = 10^7$ (second row). The energies of the initial spin configurations are shown by up and down triangles in Fig.\ \ref{SFig:BCS_E_Ana_Num_Increasing_N_Comp}(a). In the first two columns, we show the $\log_{10} \Lambda_1$ values for the spin configurations having energies close to $E_\mathrm{min}(J_0^z)$ for $J_0^z = 5$ and $J_0^z = 10$ respectively. On the other hand, the spin configurations having energies close to $E_\mathrm{max}(J_0^z)$ for $J_0^z = 5$ and $J_0^z = 10$ are considered in the last two columns. In all the figures, we denote the $\log_{10} \Lambda_1$ values for the spin configurations in the middle of the spectrum corresponding to $E(J_0^z) \approx 0$ by thick dashed lines. The uncertainties in these $\log_{10} \Lambda_1$ values are shown by thin dashed lines. The maximum Lyapunov exponent \(\Lambda_1\) shows dependence on initial spin configurations and \(J_0^z\) in (a) -- (d). Unlike this, we observe ergodic behavior for \(\tau = 10^7\) in the second row: (e) -- (h). }
    \label{SFig:No_Prethermalization_Large_tau}
\end{figure*} 

Recall that our symplectic integration scheme still conserves \(J_0^{z}\). Therefore, we analyze the dependence of the renormalized Lyapunov spectrum (LS) and the individual Lyapunov characteristic exponents (LCEs) on \(J_0^{z}\). To do this, we calculate the LS by choosing initial spin configurations that reside in different parts of the \(E(J_0^{z})\) spectrum of Fig.\ \ref{SFig:BCS_E_Ana_Num_Increasing_N_Comp}.

In Fig.\ \ref{SFig:3D_MLE}, we use spin configurations with randomly chosen $2 J^{z}_0$ spins pointing in the positive or negative $z$-direction and the rest of the spins pointing in random directions on the $xy$-plane.
These states, denoted as green crosses in Fig.\ \ref{SFig:BCS_E_Ana_Num_Increasing_N_Comp}(a), lie in the middle of the \(E(J_0^{z})\) spectrum. Completely random spin configurations with \(E \approx 0, J_0^{z} \approx 0,\) and \(\Delta \approx \mathcal{O}(1/\sqrt{N})\), which were used for the numerics of Figs.\ (1), (2), and (3) to establish the Trotter transition, also lie in the middle of this \(E(J_0^{z})\). 

Figure \ref{SFig:MLE_Reflection} displays \(\Lambda_1\) for four selected \(\tau\) -- two from the LRN regime and two from the memoryless regime -- clearly highlighting the extent of the \(J_0^{z}\) dependence. In particular, \(\Lambda_1\) depends on \(\lvert J_0^{z} \rvert\), which manifests itself in the reflection symmetries of these plots on \(J_0^{z} = 0\). In the memoryless regime, \(\Lambda_1\) exhibits a clear linear dependence on \(\lvert J_0^{z} \rvert\), with the slope decreasing as \(\tau\) increases. This behavior indicates that the system becomes fully ergodic deep within the memoryless regime. In Fig.\ \ref{SFig:Rescaled_LS_vs_Jz}, we show the rescaled LS in the two different Totter regimes for different values of \(J_0^{z}\). Here we have taken the same initial conditions as in Figs.\ \ref{SFig:3D_MLE} and \ref{SFig:MLE_Reflection}. The \(\log_{10} \overbar{\Lambda}(\rho)\) versus \(\rho\) plot in the inset of Fig.\ \ref{SFig:Rescaled_LS_vs_Jz} depicts the super-exponential decay of the LCEs as a function of $\rho \equiv i/N$.   

In Fig.\ \ref{SFig:No_Prethermalization_Large_tau}, we consider configurations that lie close to the extremal curves of the \(E(J_0^{z})\) band. To obtain perturbed states near the energy maxima curve (blue up triangles) while keeping the corresponding $J_0^z$ values fixed, we take the two spins on the domain wall -- one pointing up and the other down -- and place them in two random directions on the $xy$-plane. To perturb energy minima states without changing their $J_0^z$ (represented by red down triangles), we rotate the $(N/2)$th and the $(N/2 + 1)$th spins by random angles about the $z$-axis. 

In Fig.\ \ref{SFig:No_Prethermalization_Large_tau}(a) -- (d), the $\Lambda_1$ values for the spin configurations near the energy boundaries are suppressed to those at the middle of the spectrum -- cf. \cite{Xiaodong_sup}. Such dependence of $\Lambda_1$ on initial spin configuration strongly suggests prethermalization \cite{Pretherm1_sup, Pretherm2_sup, Pretherm3_sup, Pretherm4_sup}. In contrast, the $\Lambda_1$ values for the spin configurations near the energy boundaries are comparable to those at the middle of the spectrum in the memoryless large $\tau$ regime -- see Fig.\ \ref{SFig:No_Prethermalization_Large_tau}(e) -- (h).

\section{The Maximum Lyapunov Exponent in the Large $\tau$ Limit}
\label{Sec:Large_tau_MLE}

\begin{figure*}
    \centering
    \includegraphics[trim={0.75cm 0.75cm 0cm 0cm},clip,scale=0.90]{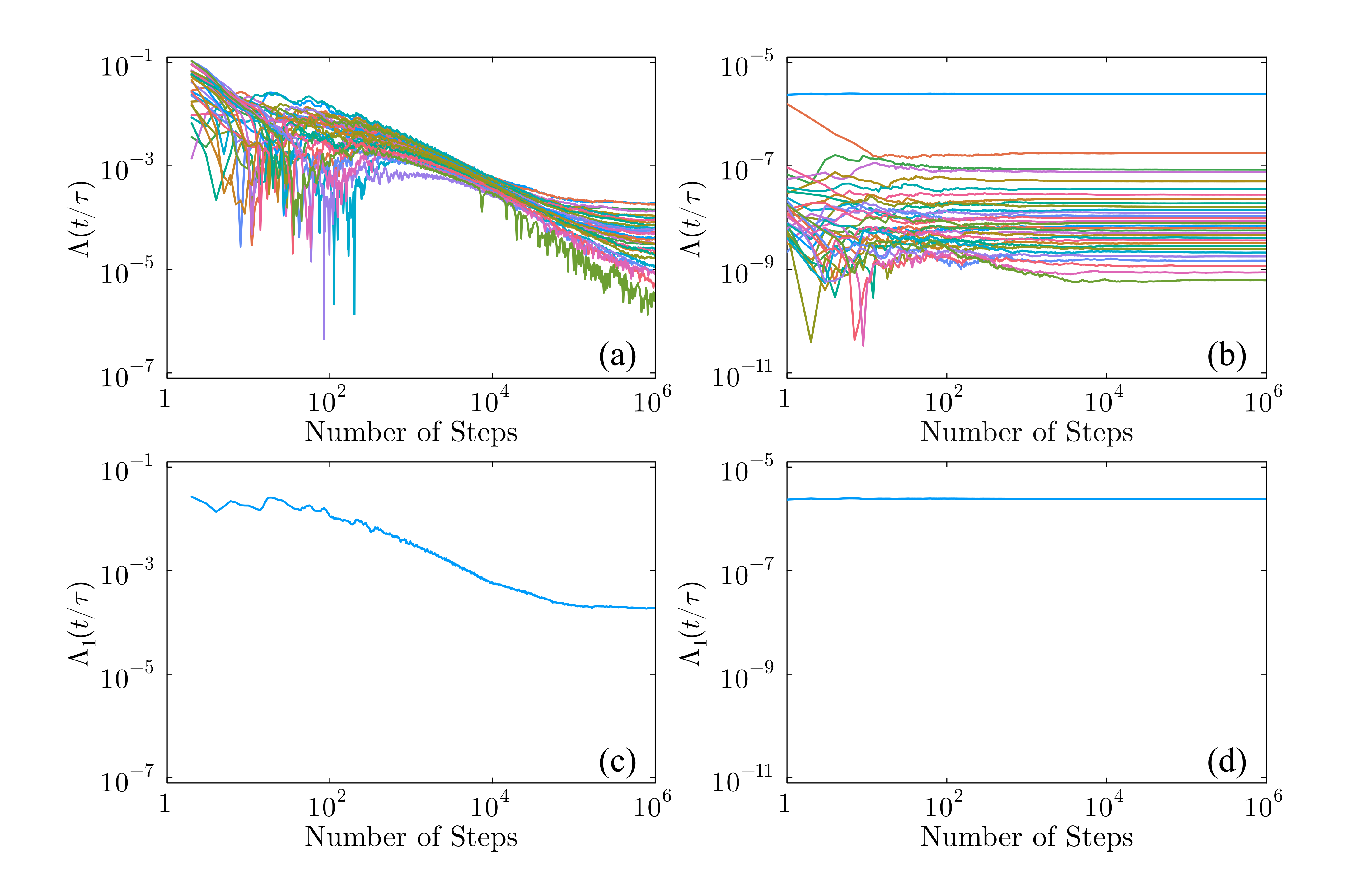}
    \caption{ Finite-time Lyapunov exponents. In panels (c) and (d), we separately plot the finite-time largest Lyapunov exponents. The Lyapunov spectrum in the first column is calculated for $\tau = 1$, whereas the one in the second column is calculated for $\tau = 10^7$. In the weakly chaotic regime, the Lyapunov exponents keep decreasing as $t^{-1}$ for long time. On the other hand, the Lyapunov exponents quickly saturate to their asymptotic values in the strongly chaotic regime. In particular, the maximum Lyapunov exponent saturates just after one time step when $\tau = 10^7$ -- see panel (d). }
    \label{SFig:Time_Dependent_LS}
\end{figure*}

In Fig.\ \ref{SFig:Time_Dependent_LS}(b), we observe that the Lyapunov exponents in the large $\tau$ regime very quickly saturate to their asymptotic values. This contrasts the small $\tau$ regime spectrum in Fig.\ \ref{SFig:Time_Dependent_LS}(a), where the exponents decay as $t^{-1}$ for a long time. This indicates that the information of the initial condition is retained for a long time in the small $\tau$ dynamics. On the other hand, the initial information is lost quickly in the large $\tau$ regime. In extreme situations, $\Lambda_1$ saturates just after one time step in Fig.\ \ref{SFig:Time_Dependent_LS}(d). This is why we denote the large $\tau$ regime as the memoryless regime. The memorylessness in the the large $\tau$ regime is also illustrated in Fig.\ \ref{SFig:No_Prethermalization_Large_tau}, where we see that in this regime one observes no initial condition dependence in $\Lambda_1$.

\subsection{Calculating $\Lambda_1$ for $N = 2$ and $J_0^z = 0$}
\label{Sec:MLE_2Spin}

We see that there is a weak $N$ dependence in the numerical values of $\Lambda_1$ in the large $\tau$ regime, see Fig.\ (6) in the main text. Therefore, we start by considering a smallest system of interacting spins with $N = 2$. In the limit of large $\tau$, the maximum Lyapunov exponent $\Lambda_1$, derived from an initial condition possessing particle-hole symmetry, is identical to that obtained from an initial condition lacking this symmetry. Therefore, it is enough to calculate $\Lambda_1$ for $N=2$ with the particle-hole symmetry, which simplifies the evolution considerably. 

Using the rules of the particle-hole symmetry in Eq.\ \eqref{SEq:PH_Def2} and suppressing the spin-index $1$, we explicitly write the single step evolution for the first spin in $N = 2$ case as follows:
\begin{equation}
    \bm{S}^\prime = e^{a_1 \tau L_\mathrm{free}}e^{b_1 \tau L_\mathrm{int}}e^{a_2 \tau L_\mathrm{free}} e^{b_1 \tau L_\mathrm{int}}e^{a_1 \tau L_\mathrm{free}} \bm{S},
\label{SEq:saba2_First_Spin_N=2}
\end{equation}
where using Eqs.\ \eqref{SEq:Tan_Map_Soln_HFree} and \eqref{SEq:Tan_Map_Soln_HInt} we obtained
\begin{equation}
    \begin{split}
    e^{L_\mathrm{free}\tau} \colon & 
    \bm{S} (\tau)     = \mathbb{R}_{\bm{z}} (  -\tau   ) \cdot \bm{S} (0),
    \\
    e^{L_\mathrm{int}\tau} \colon &
    \bm{S} (\tau)     =     \mathbb{R}_{\bm{x}} (  - 4g S^x(0) \tau   ) \cdot \bm{S} (0).
    \end{split}
\label{SEq:Trot_Steps_N=2}
\end{equation}
The evolution of the second spin is then determined using the particle-hole symmetry \eqref{SEq:PH_Def2}. In Eq.\ \eqref{SEq:Trot_Steps_N=2}, we have used $\varepsilon_1 = -1/2$. Also, as a consequence of particle-hole symmetry, we have $J^{z}_0 = 0$, $\bm{\Delta}_\mathrm{rot} = \bm{\Delta}(0) = 2gS^{x}_1 (0) \bm{x}$, and $\bm{n}_\mathrm{rot} = \bm{x}$. Note that the map in Eq.\ \eqref{SEq:saba2_First_Spin_N=2} is nothing but a generalized version of the kicked top model. In particular, applying this map several times is roughly tantamount to kicking the top twice within the same time step. Assuming \(4b_1\tau = 2\tau \gg 1\), one obtains an expression for \(\tau\Lambda_1\) as
\begin{equation}
    \label{SEq:Kicked_Top_mLCE_N=2}
    \tau\Lambda_1 \approx 2\ln\big[ 2\tau \lvert 
    \sin(2a_1\tau) \sin(a_2\tau) \rvert \big] + C_2,
\end{equation}
where \(C_2\) is a \(\tau\) independent constant \cite{KickedTop_sup}. 

\begin{figure}
    \centering
    \includegraphics[trim={0cm 0cm 0cm 0cm},clip,scale=0.04]{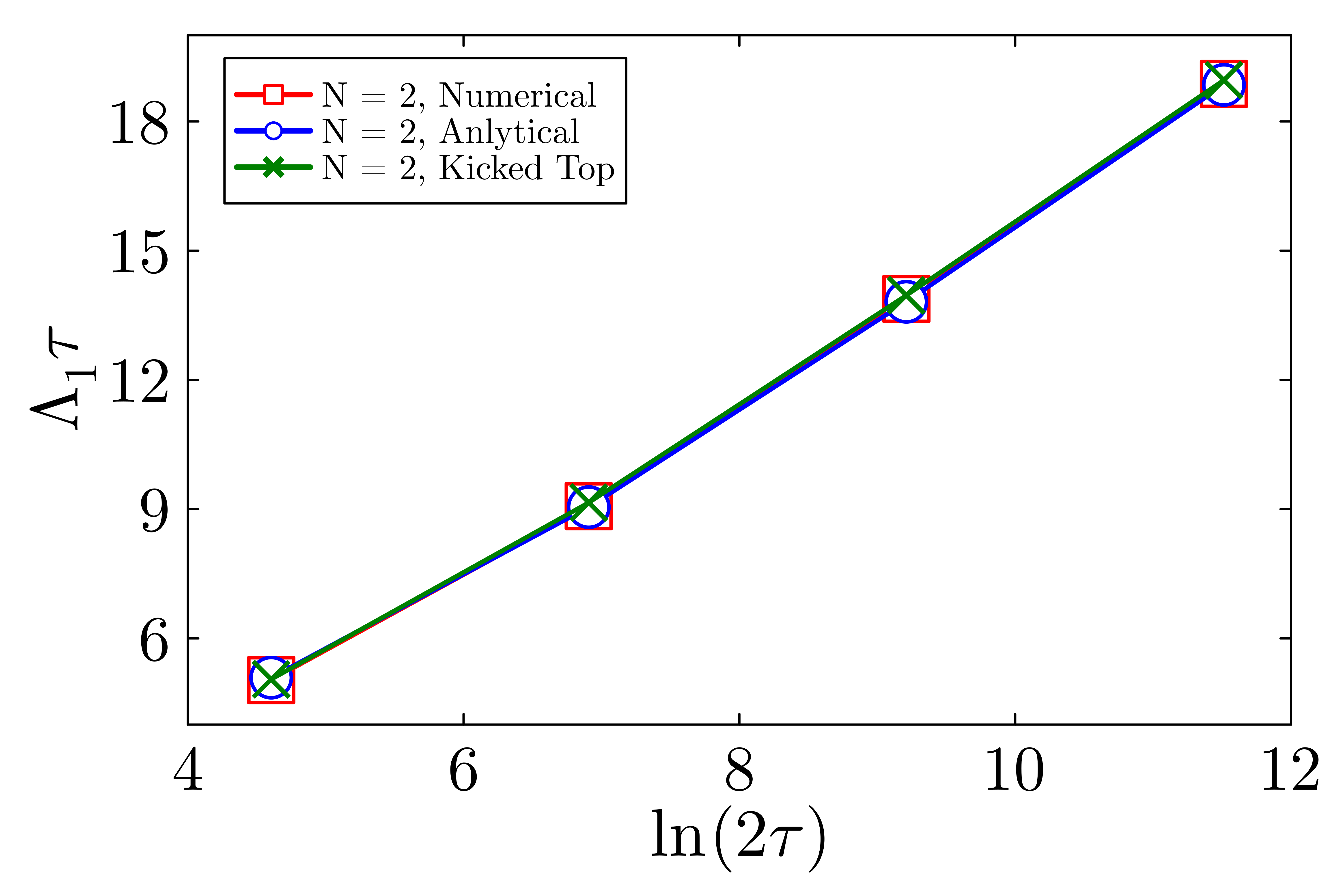}
    \caption{ We compare \(\tau \Lambda_1\) obtained from numerics (red rectangles), the analytical method of Sec.\ \ref{Sec:MLE_2Spin} using the Jacobian (blue circle), and Eq.\ \eqref{SEq:Kicked_Top_mLCE_N=2} with \(C_2 = 3.3\) (green crosses), by plotting it against \(\ln (2\tau)\). The linear fit has a slope of approximately \(2\). }
    \label{Fig:Memoryless_Scaling_N=2}
\end{figure}

Given that the finite-time largest Lyapunov exponent rapidly saturates to $\Lambda_1$ after just a single time step of evolution [see Fig.\ \ref{SFig:Time_Dependent_LS}(d)], we adopt a similar approach -- instead of directly using Eq.\ \eqref{SEq:Kicked_Top_mLCE_N=2} -- to the one used for the strongly chaotic regime of the kicked top model to compute the Lyapunov exponent in Ref.\ \cite{KickedTop_sup}. To that effect, for two spins with particle-hole symmetry, we obtain the evolved first spin after a single $\mathrm{SABA}_2$ in Eq.\ \eqref{SEq:saba2_First_Spin_N=2}, where $\bm{S}$ is a random vector on a sphere of radius $S = \lvert \bm{S} \rvert = 0.5$, which is parameterized as
\begin{equation}
\begin{split}
    S^{x} &= S \sin\theta \cos\phi, \\ 
    S^{y} &= S \sin\theta \sin\phi, \\
    S^{z} &= S \cos\theta,
\end{split}
\label{SEq:Rand_Spin}
\end{equation}
where $\cos\theta$ is chosen to be a random number between $-1$ and $+1$, and $\phi$ is a random angle chosen from the range $[0, 2\pi]$. The Jacobian matrix corresponding to the transformation \eqref{SEq:saba2_First_Spin_N=2} is written as as  
\begin{equation}
\frac{  \partial \left( S^{x^\prime}, S^{y^\prime}, S^{z^\prime} \right)  }{  \partial \left( S^{x}, S^{y}, S^{z} \right)  } = \begin{pmatrix}
\frac{  \partial S^{x^\prime}  }{  \partial S^{x}  } & \frac{  \partial S^{x^\prime}  }{  \partial S^{y}  } & \frac{  \partial S^{x^\prime}  }{  \partial S^{z}  } \\[1em]
\frac{  \partial S^{y^\prime}  }{  \partial S^{x}  } & \frac{  \partial S^{y^\prime}  }{  \partial S^{y}  } & \frac{  \partial S^{y^\prime}  }{  \partial S^{z}  } \\[1em]
\frac{  \partial S^{z^\prime}  }{  \partial S^{z}  } & \frac{  \partial S^{x^\prime}  }{  \partial S^{y}  } & \frac{  \partial S^{z^\prime}  }{  \partial S^{z}  }
\end{pmatrix}.
\label{SEq:Jac_N=2}
\end{equation} 
We then obtain the maximum Lyapunov exponent as
\begin{equation}
\Lambda_1 = \frac{1}{\tau}\left\langle \ln\left| \lambda_\textrm{max} \right| \right\rangle,  
\label{SEq:MLE_N=2}
\end{equation}
where $\lambda_\textrm{max}$ is the maximum eigenvalue of the Jacobian \eqref{SEq:Jac_N=2} and $\left\langle \right\rangle$ denotes averaging over several $\bm{S}$. In Fig.\ \ref{Fig:Memoryless_Scaling_N=2}, the numerical results for \(\tau\Lambda_1\) (red rectangles) closely match both the analytical expression from Eq.\ \eqref{SEq:MLE_N=2} (blue circles) and the approximation from Eq.\ \eqref{SEq:Kicked_Top_mLCE_N=2} with \(C_2 = 3.3\) (green crosses).

\subsection{Calculating $\Lambda_1$ for $N > 2$ and $J_0^z = 0$}
\label{Sec:MLE_NSpin}

To account for the weak $N$ dependence of $\Lambda_1$, we generalize the two-spin calculation by making $g$ in $H_\mathrm{int}$ $N$-dependent. First, we compare the relative sizes of $H_\mathrm{free}$ and $H_\mathrm{int}$ for random spin configurations. We observe that the expectation values of $H_\mathrm{free}$ and $H_\mathrm{int}$ for random spin configurations are zero. As a result, we estimate their relative sizes by their standard deviations as follows:
\begin{subequations}
\begin{equation}
\begin{split}
    \mathrm{Var}\left[ \sum_{j=1}^{N} 2\varepsilon_j S_j^{z} \right] &= \sum_{j=1}^{N} \left[ -1 + \frac{2(i-1)}{(N-1)} \right]^2 \mathrm{Var}\left(S_j^{z}\right) \\ 
    &= \bigg[ N + \frac{4}{(N-1)}\sum_{j=1}^{N}(i-1) \\
    & \qquad\qquad+ \frac{4}{(N-1)^2}\sum_{j=1}^{N}(i-1)^2 \bigg]  \mathrm{Var}\left(S_j^{z}\right) \\
    &= \left[ 3N + \frac{2N(2N-1)}{3(N-1)} \right] \mathrm{Var}\left(S_j^{z}\right), 
\end{split}
\end{equation}
\begin{equation}
    \begin{split}
    \implies H_\mathrm{free} \to \sigma_{H_\mathrm{free}} \approx \sqrt{\frac{13N}{3}} \times \mathcal{O}(1),
\end{split}  
\label{SEq:TypicalSize_H_Free}
\end{equation}
\begin{equation}
\begin{split}
    \mathrm{Var}\left( \Delta_x \right) &= \mathrm{Var}\left( g\sum_{j=1}^{N} S_j^{x} \right) \\ 
    &= g^2 \sum_{j=1}^{N} \mathrm{Var}\left( S_j^{x} \right) \approx \frac{N}{(N-1)^2} \times \mathcal{O}(1), 
\end{split}
\end{equation}
\begin{equation}
\begin{split}
    \implies H_\mathrm{int} \to \frac{\sigma_{\Delta}^2}{g} &\approx (N-1) \times \left[\frac{\sqrt{N}}{(N-1)}\right]^2\mathcal{O}(1)\\
    &\approx \mathcal{O}(1).
\end{split}
\label{SEq:TypicalSize_H_Int}
\end{equation}
\label{SEq:TypicalSize_H_Free_H_Int}
\end{subequations}
Here we have used 
\begin{equation}
    \mathrm{Var}\left( S_j^{x} \right) = \mathrm{Var}\left( S_j^{y} \right) = \mathrm{Var}\left( S_j^{z} \right) = \mathcal{O}(1)
\end{equation}
for any $j$. 

\begin{figure}[h!tb]
    \centering
    \includegraphics[trim={0.0cm 0.0cm 0.0cm 0.0cm},clip,scale=0.125]{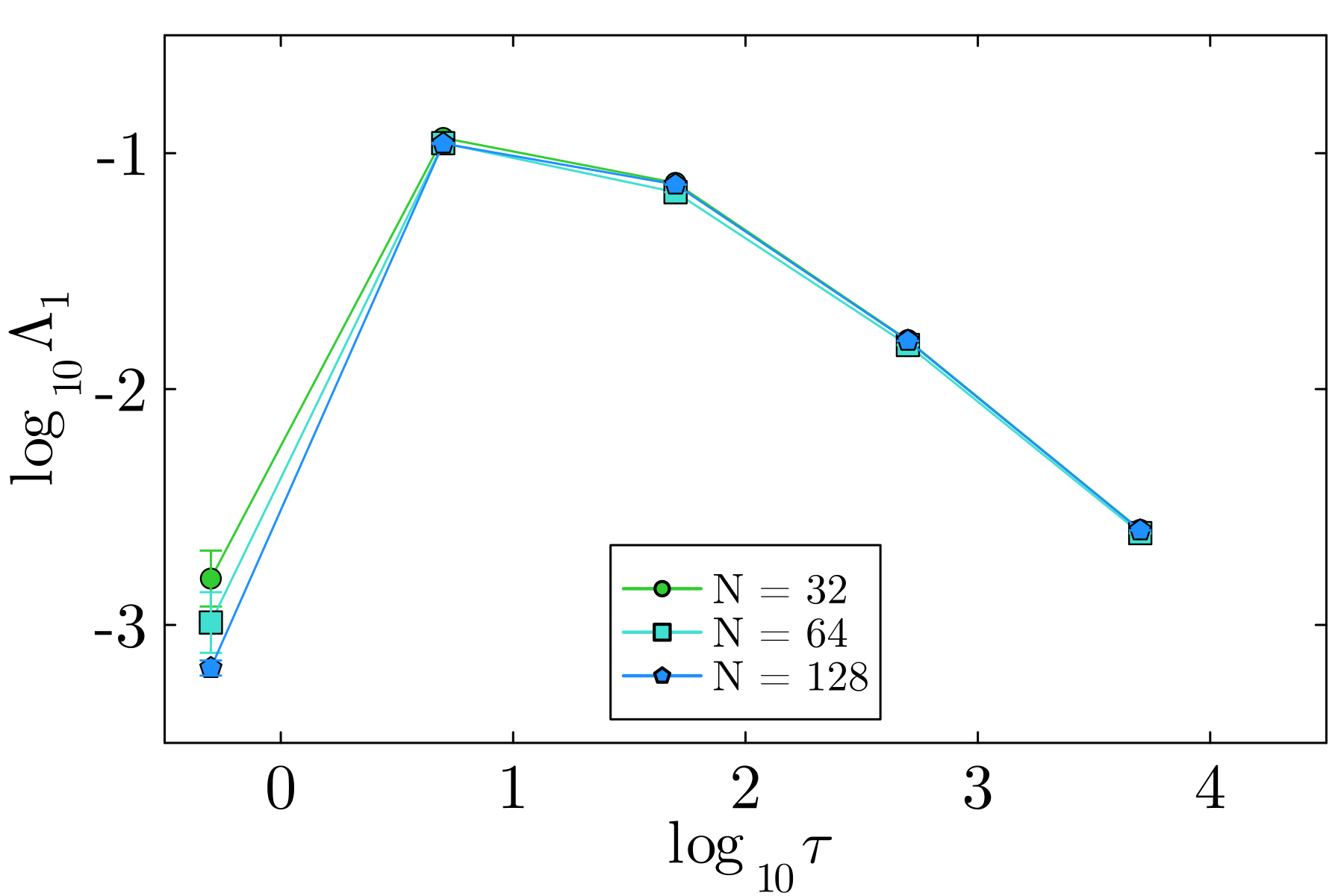}
    \caption{ The $\log$-$\log$ plot of the maximum Lyapunov exponent $\Lambda_1$ versus the step size $\tau$ for the rescaled BCS Hamiltonian with $g_\mathrm{r} = \delta \sqrt{N},$ where $\delta$ is the single-particle level spacing. We show the data for $N = 32, 64,$ and $128$ and for $0.5 \leq \tau \leq 5 \times 10^3$. We note that in the small $\tau$ regime, the differences in $\Lambda_1$ values across various $N$ have decreased significantly -- cf. Fig.\ (1) of the main text. }
    \label{SFig:Rescaled_BCS_MLE}
\end{figure}

Using Eqs.\ \eqref{SEq:TypicalSize_H_Free} and \eqref{SEq:TypicalSize_H_Int}, we observe that the typical size of $H_\mathrm{free}/H_\mathrm{int}$ is $\sqrt{13N/3}$. This explains the reason for $\Lambda_1$ values being smaller for larger $N$ values. In particular, this effect is quite pronounced in the weakly chaotic small $\tau$ regime -- see Figs.\ (1) and (6) of the main text for the small and large $\tau$ data respectively. We also observe in Fig.\ \ref{SFig:Rescaled_BCS_MLE} that scaling $H_\mathrm{int}$ with $g_\mathrm{r} = \delta\sqrt{N}$ effectively eliminates this $N$-dependence in Trotter chaos.    

We note that $H_\mathrm{free} \approx H_\mathrm{int}$ and $g = \delta = 1$ for $N = 2$. Therefore, we repeat the calculation of Sec.\ \ref{Sec:MLE_2Spin} with $g_{N} = 1/\sqrt{13N/3}$ to obtain $\Lambda_1$ for $N$-spin evolution.

\begin{thebibliography}{999}

\bibitem{Basic_Num1} D.W. Heermann, \textit{Computer-Simulation Methods}, Springer-Verlag, Berlin (1990).

\bibitem{Basic_Num2} M. Abramowitz and I. Stegun, \textit{Handbook of Mathematical Functions, With Formulas, Graphs, and Mathematical Tables}, Dover, New York (1964).




\bibitem{DQS_Feynman} R. P. Feynman, Simulating physics with computers, \href{https://doi.org/10.1007/BF02650179}{Int. J. Theor. Phys. \textbf{21}, 467 (1982)}.

\bibitem{DQS_Lloyd} S. Lloyd, Universal Quantum Simulators, \href{https://doi.org/10.1126/science.273.5278.1073}{Science \textbf{273}, 1073 (1996)}.

\bibitem{DQS_Nori} I. M. Georgescu, S. Ashhab, and F. Nori, Quantum simulation, \href{https://doi.org/10.1103/RevModPhys.86.153}{Rev. Mod. Phys. \textbf{86}, 153 (2014)}.

\bibitem{DQS_QuantH2} R. Haghshenas, E. Chertkov, M. Mills, W. Kadow, S.-H. Lin, Y. H. Chen, C. Cade, I. Niesen, Begu\u{s}i\'{c}, M. S. Rudolph, \textit{et al.}, Digital quantum magnetism at the frontier of classical simulations, \href{https://doi.org/10.48550/arXiv.2503.20870}{arXiv:2503.20870}.





\bibitem{Trot} H. F. Trotter, General theory of fractal path integrals with applications to many-body theories and statistical physics, \href{https://doi.org/10.1090/S0002-9939-1959-0108732-6}{Proc. Am. Math. Soc. \textbf{10}, 545 (1959)}.

\bibitem{Trot_Sympl_1} M. Suzuki, General theory of fractal path integrals with applications to many-body theories and statistical physics, \href{https://doi.org/10.1063/1.529425}{J. Math. Phys. \textbf{32}, 400 (1991)}.

\bibitem{Trot_Sympl_2} M. Suzuki, General theory of higher-order decomposition of exponential operators and symplectic integrators, \href{https://doi.org/10.1016/0375-9601(92)90335-J}{Phys. Lett. A \textbf{165}, 387 (1992)}.

\bibitem{Trot2} D. W. Berry, G. Ahokas, R. Cleve, and B. C. Sanders, Efficient Quantum Algorithms for Simulating Sparse Hamiltonians, \href{https://doi.org/10.1007/s00220-006-0150-x}{Commun. Math. Phys. \textbf{270}, 359 (2007)}.

\bibitem{Trot3} D. Poulin, M. B. Hastings, D. Wecker, N.Wiebe, A. C. Doherty, and M. Troyer, The trotter step size required for accurate quantum simulation of quantum Chemistry, \href{https://dl.acm.org/doi/10.5555/2871401.2871402}{Quantum Info. Comput. \textbf{15}, 361 (2015)}.

\bibitem{Trot4} R. Babbush, D. W. Berry, I. D. Kivlichan, A. Y. Wei, P. J. Love, and A. Aspuru-Guzik, Exponentially more precise quantum simulation of fermions in second quantization, \href{https://doi.org/10.1088/1367-2630/18/3/033032}{New J. Phys. \textbf{18}, 033032 (2016)}.

\bibitem{Trot5} I. Pitsios, L. Banchi, A. S. Rab, M. Bentivegna, D. Caprara, A. Crespi, N. Spagnolo, S. Bose, P. Mataloni, and R. Osellame, Photonic simulation of entanglement growth and engineering after a spin chain quench, \href{https://doi.org/10.1038/s41467-017-01589-y}{Nat. Commun. \textbf{8}, 1569 (2017)}.

\bibitem{Trot6} A. Tranter, P. J. Love, F. Mintert, N. Wiebe, and P. V. Coveney, Ordering of Trotterization: Impact on Errors in Quantum Simulation of Electronic Structure, \href{https://doi.org/10.3390/e21121218}{Entropy \textbf{21}, 1218 (2019)}.

\bibitem{Trot7} C. Cirstoiu, Z. Holmes, J. Iosue, L. Cincio, P. J. Coles, and A. Sornborger, Variational fast forwarding for quantum simulation beyond the coherence time, \href{https://doi.org/10.1038/s41534-020-00302-0}{Npj Quantum Inf. \textbf{6}, 82 (2020)}.

\bibitem{Trot8} A. Bolens and M. Heyl, Reinforcement learning for digital quantum simulation, \href{https://doi.org/10.1103/PhysRevLett.127.110502}{Phys. Rev. Lett. \textbf{127}, 110502 (2021)}.

\bibitem{Trot9} S.-H. Lin, R. Dilip, A. G. Green, A. Smith, and F. Pollmann, Real- and Imaginary-Time Evolution with Compressed Quantum Circuits, \href{https://doi.org/10.1103/PRXQuantum.2.010342}{PRX Quantum \textbf{2}, 010342 (2021)}.

\bibitem{Trot10} J. Richter and A. Pal, Simulating Hydrodynamics on Noisy Intermediate-Scale Quantum Devices with Random Circuits, \href{https://doi.org/10.1103/PhysRevLett.126.230501}{Phys. Rev. Lett. \textbf{126}, 230501 (2021)}.

\bibitem{Trot11} M. S. Tepaske, D. Hahn, and D. J. Luitz, Optimal compression of quantum many-body time evolution operators into brickwall circuits, \href{https://doi.org/10.21468/SciPostPhys.14.4.073}{SciPost Phys. \textbf{14}, 073 (2023)}.

\bibitem{Trot12} H. Zhao, M. Bukov, M. Heyl, and R. Moessner, Making Trotterization Adaptive and Energy-Self-Correcting for NISQ Devices and Beyond, \href{https://doi.org/10.1103/PRXQuantum.4.030319}{PRX Quantum \textbf{4}, 030319 (2023)}.





\bibitem{TrotterTrans1} T. Ishii, T. Kuwahara, T. Mori, and N. Hatano, Heating in Integrable Time-Periodic Systems, \href{https://doi.org/10.1103/PhysRevLett.120.220602}{Phys. Rev. Lett. \textbf{120}, 220602 (2018)}.

\bibitem{TrotterTrans2} M. Heyl, P. Hauke, and P. Zoller, Quantum localization bounds Trotter errors in digital quantum simulation, \href{https://www.science.org/doi/abs/10.1126/sciadv.aau8342}{Sci. Adv. \textbf{5}, eaau8342 (2019)}.

\bibitem{TrotterTrans3} L. M. Sieberer, T. Olsacher, A. Elben, M. Heyl, P. Hauke, F. Haake, and P. Zoller, Digital quantum simulation, Trotter errors, and quantum chaos of the kicked top, \href{https://doi.org/10.1038/s41534-019-0192-5}{npj Quantum Inf \textbf{5}, 78 (2019)}.

\bibitem{TrotterTrans4} C. Kargi, J. P. Dehollain, F. Henriques, L. M. Sieberer, T. Olsacher, P. Hauke, M. Heyl, P. Zoller, and N. K. Langford, Quantum Chaos and Universal Trotterisation Performance Behaviours in Digital Quantum Simulation, \href{https://opg.optica.org/abstract.cfm?URI=QIM-2021-W3A.1}{in F. Sciarrino, N. Treps, M. Giustina, and C. Silberhorn (eds), \textit{Quantum Information and Measurement VI 2021}, Technical Digest Series (Optica Publishing Group, 2021), paper W3A.1}.

\bibitem{TrotterTrans5} K. Chinni, M. H. Mu\~{n}oz-Arias, I. H. Deutsch, and P. M. Poggi, Trotter Errors from Dynamical Structural Instabilities of Floquet Maps in Quantum Simulation, \href{https://doi.org/10.1103/PRXQuantum.3.010351}{PRX Quantum \textbf{3}, 010351 (2022)}.

\bibitem{TrotterTrans6} E. Vernier, B. Bertini, G. Giudici, and L. Piroli, Integrable Digital Quantum Simulation: Generalized Gibbs Ensembles and Trotter Transitions, \href{https://doi.org/10.1103/PhysRevLett.130.260401}{Phys. Rev. Lett. \textbf{130}, 260401 (2023)}.

\bibitem{TrotterTrans7} T. N. Ikeda, S. Sugiura, and A. Polkovnikov, Robust Effective Ground State in a Nonintegrable Floquet Quantum Circuit, \href{https://doi.org/10.1103/PhysRevLett.133.030401}{Phys. Rev. Lett. \textbf{133}, 030401 (2024)}.

\bibitem{TrotterTrans8} P. Suchsland, R. Moessner,  and P. W. Claeys, Krylov complexity and Trotter transitions in unitary circuit dynamics, \href{https://doi.org/10.1103/PhysRevB.111.014309}{Phys. Rev. B \textbf{111}, 014309 (2025)}.

\bibitem{TrotterTrans9} P. M. Schindler and M. Bukov, Geometric Floquet Theory, \href{https://doi.org/10.1103/7l91-gw77}{Phys. Rev. X \textbf{15}, 031037 (2025)}.

\bibitem{AntiTrotterTrans} M. \v{Z}nidari\v{c}, Prethermalization, shadowing breakdown, and the absence of Trotterization transition in quantum circuits, \href{https://doi.org/10.48550/arXiv.2505.15521}{	arXiv:2505.15521 (2025)}.





\bibitem{Benettin} G. Benettin, L. Galgani, A. Giorgilli, and J.-M. Strelcyn, Lyapunov characteristic exponents for smooth dynamical systems and for Hamiltonian systems; a method for computing all of them. Part 1: Theory, \href{https://doi.org/10.1007/BF02128236}{Meccanica  \textbf{15}, 9 (1980)}.

\bibitem{Skokos1} C. Skokos and E. Gerlach, Numerical integration of variational equations, \href{https://doi.org/10.1103/PhysRevE.82.036704}{Phys. Rev. E  \textbf{82}, 036704 (2010)}.

\bibitem{CSLS} C. Skokos, The Lyapunov Characteristic Exponents and Their Computation, in J. Souchay and R. Dvorak (eds), \href{ https://doi.org/10.1007/978-3-642-04458-8_2}{Lect. Notes Phys. \textbf{790}, 63 (2010)}.

\bibitem{Haenggi} R. Eichhorn, S. J. Linz, and P. H\"{a}nggi, Transformation invariance of Lyapunov exponents, \href{https://doi.org/10.1016/S0960-0779(00)00120-X}{Chaos, Solitons \& Fractals  \textbf{12}, 1377 (2001)}.

\bibitem{Khinchin} A. Ya. Khunchin, \textit{Mathematical Foundations of Statistical Mechanics}, Dover, New York (1949).

\bibitem{Baldovin} M. Baldovin, A. Vulpiani, and G. Gradenigo, Statistical mechanics of an integrable system, \href{https://doi.org/10.1007/s10955-021-02781-7}{J. Stat. Phys.  \textbf{183}, 41 (2021)}.





\bibitem{BCS} J. Bardeen, L. N. Cooper, and  J. R. Schrieffer, Theory of Superconductivity, \href{https://doi.org/10.1103/PhysRev.108.1175}{Phys. Rev. \textbf{108}, 1175 (1957)}.

\bibitem{BCS_Anderson1} P. W. Anderson, Random-Phase Approximation in the Theory of Superconductivity, \href{https://doi.org/10.1103/PhysRev.112.1900}{Phys. Rev. \textbf{112}, 1900 (1958)}.

\bibitem{BCS_Anderson2} P. W. Anderson, Theory of dirty superconductors, \href{https://doi.org/10.1016/0022-3697(59)90036-8}{J. Phys. Chem. Solids \textbf{11}, 26 (1959)}.

\bibitem{BCS_Richardson} R. W. Richardson, Pairing in the limit of a large number of particles, \href{https://doi.org/10.1063/1.523493}{J. Math. Phys. \textbf{18}, 1802 (1977)}.

\bibitem{BCSUnivHam} I. L. Kurland, I. L. Aleiner, and B. L. Altshuler, Mesoscopic magnetization fluctuations for metallic grains close to the Stoner instability, \href{https://doi.org/10.1103/PhysRevB.62.14886}{Phys. Rev. B \textbf{62}, 14886 (2000)}.

\bibitem{BCSMFCorr} E. A. Yuzbashyan, A. A. Baytin, and B. L. Altshuler, Finite-size corrections for the pairing Hamiltonian, \href{https://doi.org/10.1103/PhysRevB.71.094505}{Phys. Rev. B \textbf{71}, 094505 (2005)}.

\bibitem{NonEqBCS1} E. A. Yuzbashyan, B. L. Altshuler, V. B. Kuznetsov, and V. Z. Enolskii, Solution for the dynamics of the BCS and central spin problems, \href{https://iopscience.iop.org/article/10.1088/0305-4470/38/36/003/meta}{J. Phys. A \textbf{38}, 7831 (2005)}.

\bibitem{NonEqBCS2} E. A. Yuzbashyan, B. L. Altshuler, V. B. Kuznetsov, and V. Z. Enolskii, Nonequilibrium Cooper pairing in the nonadiabatic regime, \href{https://doi.org/10.1103/PhysRevB.72.220503}{Phys. Rev. B \textbf{72}, 220503(R) (2005)}.

\bibitem{NonEqBCS3} E. A. Yuzbashyan, O. Tsyplyatyev and B. Altshuler, Relaxation and persistent oscillations of the order parameter in fermionic condensates, \href{https://doi.org/10.1103/PhysRevLett.96.097005}{Phys. Rev. Lett. \textbf{96}, 097005 (2006)}.

\bibitem{NonEqBCS4} A. Zabalo, A.-K. Wu, J. H. Pixley, and E. A. Yuzbashyan, Nonlocality as the source of purely quantum dynamics of BCS superconductors, \href{https://doi.org/10.1103/PhysRevB.106.104513}{Phys. Rev. B \textbf{106}, 104513 (2022)}.

\bibitem{NonEqBCS5} S. Smale, P. He, B. A. Olsen, K. G. Jackson, H. Sharum, S. Trotzky, J. Marino, A. M. Rey, and J. H. Thywissen, Observation of a transition between dynamical phases in a quantum degenerate Fermi gas, \href{https://www.science.org/doi/abs/10.1126/sciadv.aax1568}{Sci. Adv. \textbf{5}, eaax1568 (2019)}.

\bibitem{NonEqBCS6} A. Shankar, E. A. Yuzbashyan,  V. Gurarie,  P. Zoller,  J. J. Bollinger and A. M. Rey, Simulating Dynamical Phases of Chiral \( p + ip \) Superconductors with a Trapped ion Magnet, \href{https://doi.org/10.1103/PRXQuantum.3.040324}{PRX Quantum \textbf{3}, 040324 (2022)}.





\bibitem{Neri} F. Neri, \textit{Lie Algebras and Canonical Integration}, Dept. of Physics, University of Maryland, preprint (1988).

\bibitem{Yoshida} H. Yoshida, Construction of higher order symplectic integrators, \href{https://doi.org/10.1016/0375-9601(90)90092-3}{Phys. Lett. A  \textbf{150}, 262 (1990)}.

\bibitem{Koseleff1} P.-V. Koseleff, Relations among Lie formal series and construction of symplectic integrators, in G. Cohen, T. Mora, O. Moreno (eds), \textit{Applied Algebra, Algebraic Algorithms and Error-Correcting Codes, Lecture Notes (San Juan, P. R. 1993), Lecture Notes in Computer Science}, Springer, Berlin, Heidelberg (1993).

\bibitem{McLachlan1} R. I. McLachlan, Composition methods in the presence of small parameters, \href{https://doi.org/10.1007/BF01737165}{BIT Numer. Math. \textbf{35}, 258 (1995)}.

\bibitem{Koseleff2} P.-V. Koseleff, Exhaustive search of symplectic integrators using computer algebra, \href{https://bookstore.ams.org/FIC/10}{Integration algorithms and classical mechanics, Fields Inst. Commun. \textbf{10}, 103 (1996)}.

\bibitem{McLachlan2}  R. I. McLachlan, G. R. W. Quispel, and G. S. Turner, Numerical Integrators that Preserve Symmetries and Reversing Symmetries, \href{https://doi.org/10.1137/S0036142995295807}{SIAM J. Numer. Anal. \textbf{35}, 586 (1998)}.

\bibitem{Laskar} J. Laskar and P. Robutel, High order symplectic integrators for perturbed Hamiltonian systems, \href{https://doi.org/10.1023/A:1012098603882}{Celest. Mech. Dyn. Astron.  \textbf{80}, 39 (2001)}.

\bibitem{Tao} M. Tao, Explicit symplectic approximation of nonseparable Hamiltonians: Algorithm and long time performance, \href{https://doi.org/10.1103/PhysRevE.94.043303}{Phys. Lett. E  \textbf{94}, 043303 (2016)}.

\bibitem{Skokos2} C. Danieli, B. M. Manda, T. Mithun, and C. Skokos, Computational efficiency of numerical integration methods for the tangent dynamics of many-body Hamiltonian systems in one and two spatial dimensions, \href{https://www.aimspress.com/article/10.3934/mine.2019.3.447}{Math. Eng.  \textbf{1}, 447 (2019)}.

\bibitem{AP} C. Danieli, E. A. Yuzbashyan, B. L. Altshuler, A. Patra, and S. Flach, Dynamical chaos in the integrable Toda chain induced by time discretization, \href{https://doi.org/10.1063/5.0171261}{Chaos \textbf{34}, 033107 (2024)}.





\bibitem{FloquetHeat_Rigol} L. D’Alessio and M. Rigol, Long-time Behavior of Isolated Periodically Driven Interacting Lattice Systems, \href{https://doi.org/10.1103/PhysRevX.4.041048}{Phys. Rev. X \textbf{4}, 041048 (2014)}.

\bibitem{FloquetHeat_Lazarides} A. Lazarides, A. Das, and R. Moessner, Equilibrium states of generic quantum systems subject to periodic driving, \href{https://doi.org/10.1103/PhysRevX.4.041048}{Phys. Rev. E \textbf{90}, 012110 (2014)}.

\bibitem{FloquetHeat_Papic} P. Ponte, A. Chandran, Z. Papi\'{c}, and D. A. Abanin, Periodically driven
ergodic and many-body localized quantum systems, \href{https://doi.org/10.1016/j.aop.2014.11.008}{Ann. Phys. \textbf{353}, 196 (2015)}.

\bibitem{FloquetHeat_Abanin1} D. A. Abanin, W. De Roeck, and F. Huveneers, Exponentially Slow Heating in Periodically Driven Many-Body Systems, \href{https://doi.org/10.1103/PhysRevLett.115.256803}{Phys. Rev. Lett. \textbf{115}, 256803 (2015)}.

\bibitem{FloquetHeat_Mori1} T. Mori, T. Kuwahara, and K. Saito, Rigorous Bound on Energy Absorption and Generic Relaxation in Periodically Driven Quantum Systems, \href{https://doi.org/10.1103/PhysRevLett.116.120401}{Phys. Rev. Lett. \textbf{116}, 120401 (2016)}.

\bibitem{FloquetHeat_Mori2} T. Mori, T. Kuwahara, and K. Saito, Floquet–Magnus theory and generic transient dynamics in periodically driven many-body quantum systems, \href{https://doi.org/10.1016/j.aop.2016.01.012}{Ann. Phys. \textbf{367}, 96 (2016)}.

\bibitem{FloquetHeat_Abanin2} D. Abanin, W. De Roeck, W.W. Ho, and F. Huveneers, Effective Hamiltonians, prethermalization, and slow energy absorption in periodically driven many-body systems, \href{https://doi.org/10.1103/PhysRevB.95.014112}{Phys. Rev. B \textbf{95}, 014112 (2017)}.

\bibitem{FloquetHeat_Abanin3} D. A. Abanin, W. De Roeck, and F. Huveneers, A Rigorous Theory of Many-Body Prethermalization for Periodically Driven and Closed Quantum Systems, \href{https://doi.org/10.1007/s00220-017-2930-x}{Commun. Math. Phys. \textbf{354}, 809 (2017)}.

\bibitem{FloquetHeat_Luitz} D. J. Luitz, Y. B. Lev, and A. Lazarides, Absence of dynamical localization in interacting driven systems, \href{https://doi.org/10.21468/SciPostPhys.3.4.029}{SciPost Phys. \textbf{3}, 029 (2017))}.






\bibitem{NumChaos1} M. Ablowitz, B. Herbst, and C. Schober, Computational chaos in the nonlinear Schr\"{o}dinger equation without homoclinic crossings, \href{https://doi.org/10.1016/0378-4371(95)00434-3}{Phys. A \textbf{228}, 212 (1996)}.

\bibitem{NumChaos2} A. Calini, N. Ercolani, D. McLaughlin, and C. Schober, Mel'nikov analysis of numerically induced chaos in the nonlinear Schr\"{o}dinger equation, \href{https://doi.org/10.1016/0167-2789(95)00223-5}{Phys. D \textbf{89}, 227 (1996)}.

\bibitem{NumChaos3} M. Ablowitz, B. Herbst, and C. Schober, On the numerical solution of the Sine–Gordon equation, \href{https://doi.org/10.1006/jcph.1996.5606}{J. Comput. Phys. \textbf{131}, 354 (1997)}.

\bibitem{NumChaos4} M. J. Ablowitz, Y. Ohta, and A. D. Trubatch, On integrability and chaos in
discrete systems, \href{https://doi.org/10.1016/S0960-0779(98)00280-X}{Chaos, Solitons \& Fractals  \textbf{11}, 159 (2000)}.

\bibitem{NumChaos5} M. Ablowitz, B. Herbst, and C. Schober, Discretizations, integrable systems
and computation, \href{https://doi.org/10.1088/0305-4470/34/48/330}{J. Phys. A: Math. Gen. \textbf{34}, 10671 (2001)}.

\bibitem{NumChaos6} A. Islas, D. Karpeev, and C. Schober, Geometric integrators for the nonlinear
Schr\"{o}dinger equation, \href{https://doi.org/10.1006/jcph.2001.6854}{J. Comput. Phys. \textbf{173}, 116 (2001)}.

\bibitem{NumChaos7} B. J. Sung, J. H. Moon, and M. S. Kim, Checking the influence of numerically induced chaos in the computational study of intramolecular dynamics using trajectory equivalence, \href{https://doi.org/10.1016/S0009-2614(01)00624-8}{Chem. Phys. Lett. \textbf{342}, 610 (2001)}.

\bibitem{NumChaos8} D. Triadis, P. Broadbridge, K. Kajiwara, and K. Maruno, Integrable discrete model for one-dimensional soil water infiltration, \href{https://doi.org/10.1111/sapm.12208}{Stud. Appl. Math. \textbf{140}, 483 (2018)}.






\bibitem{KickedTop_Haake} F. Haake, M. Ku\'{s}, and R. Scharf, Classical and Quantum Chaos for a Kicked Top, \href{https://doi.org/10.1007/BF01303727}{Z. Phys. B -- Condensed Matter \textbf{65}, 381 (1987)}.

\bibitem{KickedTop_Greeks} V. Constantoudis and N. Theodorakopoulos, Lyapunov exponent, stretching numbers, and islands of stability of the kicked top, \href{https://doi.org/10.1103/PhysRevE.56.5189}{Phys. Rev. E \textbf{56}, 5189 (1997)}.

\bibitem{KickedTop_Deutsch} M. H. Mu\~{n}oz-Arias, P. M. Poggi, and I. H. Deutsch, Nonlinear dynamics and quantum chaos of a family of kicked $p$-spin models, \href{https://doi.org/10.1103/PhysRevE.103.052212}{Phys. Rev. E \textbf{103}, 052212 (2021)}.

\bibitem{KickedTop_Ghose} A. Anand, R. B. Mann,  and S. Ghose, Non-linearity and chaos in the kicked top, \href{https://doi.org/10.1016/j.physd.2024.134455}{Physica D \textbf{471}, 134455 (2025)}.

\bibitem{LMG} H.J. Lipkin, N. Meshkov, and A.J. Glick, Validity of many-body approximation methods for a solvable model: (I). Exact solutions and perturbation theory, \href{https://doi.org/10.1016/0029-5582(65)90862-X}{Nucl. Phys. \textbf{62}, 188 (1965)}.




\bibitem{KAM11} A. A. N. Kolmogorov, Preservation of conditionally periodic movements with small change in the Hamilton function, in G. Casati and J. Ford (eds), \textit{Stochastic Behavior in Classical and Quantum Hamiltonian Systems, Lecture Notes in Physics, vol 93}, Springer, Berlin, Heidelberg (1979).

\bibitem{KAM12} V. I. Arnol'd, \textit{Mathematical methods of classical mechanics}, Springer, New York, NY (2013).

\bibitem{KAM13} J. Moser, On invariant curves of area-preserving mapping of an annulus, \href{https://www.mathnet.ru/eng/mat236}{Matematika \textbf{6}, 51 (1962)}.

\bibitem{KAM2} A. M. O. De Almeida, \textit{Hamiltonian systems: Chaos and quantization}, Cambridge University Press, Cambridge (1988).

\bibitem{KAM3} J. P\"{o}schel, A lecture on the classical KAM theorem, \href{https://doi.org/10.1090/pspum/069/1858551}{Proc. Symp. Pure Math \textbf{69}, 707 (2000)}.

\bibitem{KAM4} C. E. Wayne, The KAM theory of systems with short range interactions, I, \href{https://doi.org/10.1007/BF01214577}{Comm. Math. Phys. \textbf{96}, 311 (1984)}.

\bibitem{KAM_Perturb1} L. Chierchia and G. Gallavotti, Smooth prime integrals for quasi-integrable Hamiltonian systems, \href{https://doi.org/10.1007/BF02721167}{Nuov. Cim. B \textbf{67}, 277 (1982)}.

\bibitem{KAM_Perturb2} G. Benettin, L. Galgani, A. Giorgilli, and J.-M. Strelcyn, A proof of Kolmogorov’s theorem on invariant tori using canonical transformations defined by the Lie method, \href{https://doi.org/10.1007/BF02748972}{Nuov. Cim. B \textbf{79}, 201 (1984)}.

\bibitem{KAM_Perturb3} M. Falcioni, U. Marini Bettolo Marconi, and A. Vulpiani, Ergodic properties of high-dimensional symplectic maps, \href{https://doi.org/10.1103/PhysRevA.44.2263}{Phys. Rev. A \textbf{44}, 2263 (1991)}.





\bibitem{ClassHeisen} V. Constantoudis and N. Theodorakopoulos, Nonlinear dynamics of classical Heisenberg chains, \href{https://doi.org/10.1103/PhysRevE.55.7612}{Phys. Rev. E \textbf{55}, 7612 (1997)}.

\bibitem{Mithun1} T. Mithun, Y. Kati, C. Danieli, and S. Flach, Weakly Nonergodic Dynamics in the Gross-Pitaevskii Lattice, \href{https://doi.org/10.1103/PhysRevLett.120.184101}{Phys. Rev. E \textbf{120}, 184101 (2018)}.

\bibitem{CarloGlass} C. Danieli, T. Mithun, Y. Kati, D. K. Campbell, and S. Flach, Dynamical glass in weakly nonintegrable Klein-Gordon chains, \href{https://doi.org/10.1103/PhysRevE.100.032217}{Phys. Rev. E \textbf{100}, 032217 (2019)}.

\bibitem{Mithun2} T. Mithun, C. Danieli, Y. Kati, and S. Flach, Dynamical Glass and Ergodization Times in Classical Josephson Junction Chains, \href{https://doi.org/10.1103/PhysRevLett.122.054102}{Phys. Rev. Lett. \textbf{122}, 054102 (2019)}.

\bibitem{Lubini} S. Iubini and A. Politi, Chaos and localization in the discrete nonlinear Schr\"{o}dinger equation, \href{https://doi.org/10.1016/j.chaos.2021.110954}{Chaos, Solitons \& Fractals  \textbf{147}, 110954 (2021)}.

\bibitem{Mithun3} T. Mithun, C. Danieli, M. V. Fistul, B. L. Altshuler, and S. Flach, Fragile many-body ergodicity from action diffusion, \href{https://doi.org/10.1103/PhysRevE.104.014218}{Phys. Rev. E \textbf{104}, 014218 (2022)}.

\bibitem{MerabPRL} M. Malishava and S. Flach, Lyapunov Spectrum Scaling for Classical Many-Body Dynamics Close to Integrability, \href{https://doi.org/10.1103/PhysRevLett.128.134102}{Phys. Rev. Lett. \textbf{128}, 134102 (2022)}; Erratum \href{https://doi.org/10.1103/PhysRevLett.130.199901}{Phys. Rev. Lett. \textbf{130}, 199901 (2023)}.

\bibitem{MerabChaos} M. Malishava and S. Flach, Thermalization dynamics of macroscopic weakly nonintegrable maps, \href{https://doi.org/10.1063/5.0092032}{Chaos \textbf{32}, 063113 (2022)}.

\bibitem{Gabriel} G. M. Lando and S. Flach, Thermalization slowing down in multidimensional Josephson junction networks, \href{https://doi.org/10.1103/PhysRevE.108.L062301}{Phys. Rev. E \textbf{108}, L062301 (2023)}.

\bibitem{Weihua} W. Zhang, G. M. Lando, B. Dietz, and S. Flach, Thermalization universality-class transition induced by Anderson localization, \href{https://doi.org/10.1103/PhysRevResearch.6.L012064}{Phys. Rev. Research \textbf{6}, L012064 (2024)}.

\bibitem{Xiaodong} X. Zhang, G. M. Lando, B. Dietz, and S. Flach, Observation of prethermalization in weakly nonintegrable unitary maps, \href{https://pcs.ibs.re.kr/f51_0870e.pdf}{Fizyka Nyzkykh Temperatur/Low Temperature Physics \textbf{51}, 870 (2025)}.




\bibitem{Supp} Supplemental Material.


\end{thebibliography}

\begin{thebibliography}{99}



\bibitem{Yoshida_sup} H. Yoshida, Construction of higher order symplectic integrators, \href{https://doi.org/10.1016/0375-9601(90)90092-3}{Phys. Lett. A  \textbf{150}, 262 (1990)}.

\bibitem{Laskar_sup} J. Laskar and P. Robutel, High order symplectic integrators for perturbed Hamiltonian systems, \href{https://doi.org/10.1023/A:1012098603882}{Celest. Mech. Dyn. Astron.  \textbf{80}, 39 (2001)}.

\bibitem{Tao_sup} M. Tao, Explicit symplectic approximation of nonseparable Hamiltonians: Algorithm and long time performance, \href{https://doi.org/10.1103/PhysRevE.94.043303}{Phys. Lett. E  \textbf{94}, 043303 (2016)}.

\bibitem{Skokos2_sup} C. Danieli, B. M. Manda, T. Mithun, and C. Skokos, Computational efficiency of numerical integration methods for the tangent dynamics of many-body Hamiltonian systems in one and two spatial dimensions, \href{https://www.aimspress.com/article/10.3934/mine.2019.3.447}{Math. Eng.  \textbf{1}, 447 (2019)}.




\bibitem{Skokos1_sup} C. Skokos and E. Gerlach, Numerical integration of variational equations, \href{https://doi.org/10.1103/PhysRevE.82.036704}{Phys. Rev. E  \textbf{82}, 036704 (2010)}.

\bibitem{Benettin_sup} G. Benettin, L. Galgani, A. Giorgilli, and J.-M. Strelcyn, Lyapunov characteristic exponents for smooth dynamical systems and for Hamiltonian systems; a method for computing all of them. Part 1: Theory, \href{https://doi.org/10.1007/BF02128236}{Meccanica  \textbf{15}, 9 (1980)}.

\bibitem{CSLS_sup} C. Skokos, The Lyapunov Characteristic Exponents and Their Computation, in J. Souchay and R. Dvorak (eds), \href{ https://doi.org/10.1007/978-3-642-04458-8_2}{Lect. Notes Phys. \textbf{790}, 63 (2010)}.




\bibitem{AidanBCS_sup} A. Zabalo, A.-K. Wu, J. H. Pixley, and E. A. Yuzbashyan, Nonlocality as the source of purely quantum dynamics of BCS superconductors, \href{https://doi.org/10.1103/PhysRevB.106.104513}{Phys. Rev. B \textbf{106}, 104513 (2022)}.

\bibitem{Gabriel_sup} G. M. Lando and S. Flach, Thermalization slowing down in multidimensional Josephson junction networks, \href{https://doi.org/10.1103/PhysRevE.108.L062301}{Phys. Rev. E \textbf{108}, L062301 (2023)}.

\bibitem{Xiaodong_sup} X. Zhang, G. M. Lando, B. Dietz, and S. Flach, Observation of prethermalization in weakly nonintegrable unitary maps, \href{https://pcs.ibs.re.kr/f51_0870e.pdf}{Fizyka Nyzkykh Temperatur/Low Temperature Physics \textbf{51}, 870 (2025)}.

\bibitem{KickedTop_sup} V. Constantoudis and N. Theodorakopoulos, Lyapunov exponent, stretching numbers, and islands of stability of the kicked top, \href{https://doi.org/10.1103/PhysRevE.56.5189}{Phys. Rev. E \textbf{56}, 5189 (1997)}.





\bibitem{Pretherm1_sup} J. C. Halimeh, V. Zauner-Stauber, I. P. McCulloch, I. de Vega, U. Schollw\"{o}ck, and M. Kastner, Prethermalization and persistent order in the absence of a thermal phase transition, \href{https://doi.org/10.1103/PhysRevB.95.024302}{Phys. Rev. B \textbf{95}, 024302 (2017)}.

\bibitem{Pretherm2_sup} B. Neyenhuis, J. Zhang, P. W. Hess, J. Smith, A. C. Lee, P. Richerme, Z.-X. Gong, A. V. Gorshkov, and C. Monroe, Observation of prethermalization in long-range interacting spin chains, \href{https://doi.org/10.1126/sciadv.1700672}{Sci. Adv. \textbf{3}, e1700672 (2017)}.

\bibitem{Pretherm3_sup} T. Mori, Floquet prethermalization in periodically driven classical spin systems, \href{https://doi.org/10.1103/PhysRevB.98.104303}{Phys. Rev. B \textbf{98}, 104303 (2018)}.

\bibitem{Pretherm4_sup} D. J. Luitz, R. Moessner, S. L. Sondhi, and V. Khemani, Prethermalization without Temperature, \href{https://doi.org/10.1103/PhysRevX.10.021046}{Phys. Rev. X \textbf{10}, 021046 (2020)}.



\end{thebibliography}
\end{document}